\documentclass{aa} 
\usepackage{amsmath}
\usepackage[boxed]{algorithm2e}
\usepackage{algorithmic}
\usepackage{enumerate}
\usepackage{epsfig,subfigure,color,amsmath,graphicx,multirow,subfigmat,ctable,bm}
\usepackage{natbib}

\newcommand{\beq}{\begin{equation}}
\newcommand{\eeq}{\end{equation}}
\newcommand{\beqs}{\begin{eqnarray}}
\newcommand{\eeqs}{\end{eqnarray}}

\input{epsf} 
\newcommand{\bfi}{\begin{figure} \epsfxsize=8cm \epsffile}
\newcommand{\efi}{\end{figure}} 

\begin{document}

\title{Testing the Analytical Blind Separation  method in simulated CMB polarization maps}

\author{Larissa Santos \inst{1,2}, Jian Yao \inst{3}, Le Zhang \inst{3}, Shamik Ghosh \inst{4,5}, Pengjie Zhang \inst{3,6,7}, Wen Zhao \inst{4,5}, Thyrso Villela \inst{8,9}, Jiming Chen \inst{4,5}, Jacques Delabrouille \inst{10,11,5}}

\institute{Center for Gravitation and Cosmology, College of Physical Science and Technology, Yangzhou University, Yangzhou 225009, China \and School of Aeronautics and Astronautics, Shanghai Jiao Tong University, Shanghai 200240, China \and Department of Astronomy, Shanghai Jiao Tong University, Shanghai, 200240, China \and CAS Key Laboratory for Research in Galaxies and Cosmology, Department of Astronomy, University of Science and Technology of China, Hefei 230026, China \and School of Astronomy and Space Sciences, University of Science and Technology of China, Hefei, 230026, China \and IFSA Collaborative Innovation Center, Shanghai Jiao Tong University, Shanghai 200240, China \and  Tsung-Dao Lee Institute, Shanghai 200240, China  \and Divis\~ao de Astrof\'isica, Instituto Nacional de Pesquisas Espaciais (INPE), 12227-010 - S\~ao Jos\'e dos Campos, SP, Brazil \and Instituto de F\'isica, Universidade de Bras\'ilia, 70919-970 - Bras\'ilia, DF, Brazil \and Laboratoire Astroparticule et Cosmologie (APC), CNRS/IN2P3, Universit\'e Paris Diderot, 75205 Paris Cedex 13, France \and IRFU, CEA, Universit\'e Paris Saclay, 91191 Gif-sur-Yvette, France}

\offprints{L. Santos \email{larissa@yzu.edu.cn}}

\abstract{Multi-frequency observations are needed to separate the CMB from foregrounds and accurately extract cosmological information from the data. The Analytical Blind Separation (ABS) method is dedicated to extracting the CMB power spectrum from multi-frequency observations in the presence of contamination from astrophysical foreground emission and instrumental noise.}{In this study, we apply the ABS method to simulated sky maps as could be observed with a future space-borne survey, in order to test the method's capability for determining the CMB polarization $E$- and $B$-mode power spectra.} {We present the ABS method performance on simulations for both a full-sky analysis and for an analysis concentrating on sky regions less impacted by Galactic foreground emission.}{ We discuss the origin and minimization of biases in the estimated CMB polarization angular power spectra. We find that the ABS method performs quite well for the analysis of full-sky observations at intermediate and small angular scales, in spite of strong foreground contamination. On the largest scales, extra work is still required to reduce biases of various origins and the impact of confusion between CMB $E$ and $B$ polarization for partial- sky analyses.}
{}

\titlerunning{Testing the ABS method in simulated CMB polarization maps}
\authorrunning{Larissa Santos et al.}

\maketitle

\section{Introduction}

Cosmic Microwave Background (CMB) experiments detect a superposition of microwave sky emissions, in which the CMB signal is mixed with signals originating from various astrophysical foreground sources among which, in particular, diffuse emissions from the Galactic interstellar medium (ISM). Multi-frequency observations are needed to separate these different emissions on the basis of their different colors and accurately measure the CMB angular power spectra, enabling the estimation of the cosmological information of interest with adequate precision. 

In the past decades, many ground-based and balloon-borne experiments have been dedicated to CMB observations. We may cite, for instance, Boomerang~\citep{2000Natur.404..955D}, MAXIMA~\citep{2000ApJ...545L...5H}, Archeops~\citep{2003A&A...399L..19B}, DASI~\citep{2002ApJ...568...38H},  VSA~\citep{2003MNRAS.341.1057W}, CBI~\citep{2003ApJ...591..540M},  ACBAR~\citep{2004ApJ...600...32K}, BEAST~\citep{2005ApJS..158..101M,2005ApJS..158..93O}, SPT~\citep{2013JCAP...10..060S}, ACT~\citep{2014JCAP...04..014D}, POLARBEAR~\citep{2014ApJ...794..171P}, SPTpol~\citep{2015ApJ...807..151K}. The CMB was also observed from space with
Relikt~1~\citep{1992SvAL...18..153S}, COBE~\citep{1992ApJ...396L...1S,1994ApJ...420..439M}, WMAP~\citep{2003ApJS..148....1B,2013ApJS..208...19H}
and most recently with the Planck space mission~\citep{2014AA...571A..16P}. The latest results from the Planck satellite achieve a precise measurement not only of temperature anisotropies, but also of CMB polarization $E$-modes~\citep{planck-e-mode, planck-e-mode2, 2019arXiv190712875P}, which, just like the CMB temperature anisotropies, are primarily generated by scalar density perturbations in the early universe.

As an outcome of these experiments, much cosmological information has already been extracted from the CMB. This resulted in the establishment of the standard $\Lambda$-CDM cosmological model, with tight constraints on the values of its main parameters~\citep{2018arXiv180706209P}. Recently, much attention has been focused on CMB polarization anisotropies. Polarization $E$-modes, of even parity, convey information complementary to that of temperature anisotropies, allowing to lift some of the degeneracies between cosmological models and parameter sets and to confirm the consistency of the global cosmological scenario.
Polarization $B$-modes, of odd parity, are of particular interest as they are expected to probe tensor perturbations due to the primordial gravitational waves~\citep{zal+seljak:1997,kamionkowski97,Hu2002}, potentially generated during a period of cosmic inflation ~\citep{pgw1,pgw2,pgw3}. 
On small scales, CMB $B$-modes mostly originate from the gravitational lensing of $E$-modes on the path of the CMB photons from the last scattering surface to the observers~\citep{zaldarriaga-lensing,hu-lensing,lewis-review}. The level of large-scale additional $B$-modes due to inflationary gravitational waves, if any, is not known. It's determination is one of the primary objectives of upcoming CMB polarization experiments, such as WMPol~\citep{2008ApJS..177..101M},
QUBIC~\citep{battistelli/etal:2011}, 
BICEP3~\citep{2018SPIE10708E..2NK}, 
AliCPT~\citep{2017arXiv171003047L}, 
CLASS~\citep{2014SPIE.9153E..1IE}, 
Simons Observatory~\citep{2019JCAP...02..056A}, 
CMB-S4~\citep{2016arXiv161002743A}, 
as well as next-generation space missions such as
CORE~\citep{2018JCAP...04..014D}, 
LiteBIRD~\citep{2014JLTP..176..733M}, 
EPIC/CMBpol~\citep{2009arXiv0906.1188B}, 
PIXIE~\citep{2011JCAP...07..025K}, 
PRISM~\citep{2014JCAP...02..006A},
PICO~\citep{2019arXiv190210541H}. The ultimate scientific exploitation of the CMB as a cosmological probe could also be achieved with a powerful spectro-polarimetric survey of the microwave sky~\citep{2019arXiv190901591D,2019arXiv190901592B,2019arXiv190901593C}.

Foreground emissions are particularly critical for measuring CMB $B$-modes, which even on the cleanest patches of sky are fainter than polarized Galactic ISM synchrotron and dust emissions \citep{2016AA...594A..10P, planck-foregrounds,2018arXiv180104945P}.
To best control the impact of foreground contamination on experiments which aim at precisely measuring CMB polarization, there is a need for data analysis tools to separate the CMB signal from its contaminants in a reliable way and characterize any possible errors in the measured angular power spectra.

Foreground contamination is lower if one concentrates the observations of CMB $B$-modes in regions of low foreground emission, away, in particular, from the strong emission of the Galactic interstellar medium in the vicinity of the Galactic plane. However, there is a drawback to this strategy: the measurement of $B$-modes then suffers from the so-called $E$-$B$ mixing, which arises from $E$-$B$ decomposition on an incomplete sky coverage ~\citep{2001PhRvD..64f3001T}. The accuracy of the measurement of the $B$-mode angular power spectrum is thus dependent on both our ability to decompose the CMB polarization signal with a partial-sky coverage, and on the effectiveness of the component separation methods. 

In this paper, we discuss the estimation of CMB $E$-mode and $B$-mode power spectra in the presence of foreground contamination using the ABS method~\citep{2016arXiv160803707Z}. We test ABS both in a full-sky and a part sky analysis to compare its performance in these two cases.
Although an important issue for precise CMB polarization measurements, we do not discuss the impact of instrumental systematic effects, and in particular of those systematic effects that induce $T$-$E$ and $T$-$B$ mixing. We refer the reader to some relevant dedicated studies \citep{2002AIPC..609..209K,2003PhRvD..67d3004H,rosset2007,2008ApJ...689..655M,2008PhRvD..77h3003S,2013ApJ...762L..23K,2015ApJ...814..110B,2018JCAP...04..022N,2017JCAP...12..015T,2019JCAP...07..043B}.

The rest of the paper is organised as follows: Section \ref{sec:compsep} introduces the state of the art in component separation for CMB observations, and gives a short description of the ABS method and how it compares with other approaches. Section \ref{sec:biases} discusses possible sources of biases for CMB power spectrum estimation with ABS. Section \ref{sect:map} describes the generation of simulated data for testing ABS, and the analysis methodology. Results are presented in Section \ref{results}, before we conclude in Section \ref{sect:con}. Three appendixes complement this by discussing a case with an extra foreground component due to polarized anomalous microwave emission (Appendix~\ref{app:AME}); an investigation of the effect of $E$-to-$B$ leakage correction for partial-sky analysis (Appendix \ref{app:E-B}); and an analytic illustration of the origin of biases in a toy-model case (Appendix \ref{appendix:special-case}).

\section{Component separation and CMB observations}
\label{sec:compsep}

\subsection{State of the art}

Various methods exist to separate the emission of foreground astrophysical components from the CMB signal in multi-frequency CMB observations. They have been widely used for the analysis of existing observations, in particular those of the WMAP and Planck space missions. These methods rely on different assumptions about the CMB and foreground emission properties, such as frequency dependence of emission, correlation between frequencies, parametric form of emission law across frequencies and/or angular power spectrum. They also differ in their optimization criteria and practical implementation.

Some methods assume that each component is characterised by a specific frequency-dependent emission law (i.e., scaling of emission as a function of frequency). The functional form of this frequency dependence, and the value of its parameters, may or may not be assumed to be known a priori. Methods that assume prior information about the frequency dependence of foreground emission (and some assumptions about their spatial distributions) include Wiener Filtering ~\citep{1994ApJ...432L..75B,1996MNRAS.281.1297T,1999MNRAS.302..663B} and the Maximum Entropy Method~\citep{1998MNRAS.300....1H}. The Gibbs sampling approaches proposed and implemented by \cite{2004ApJ...609....1J,2004PhRvD..70h3511W,2004ApJS..155..227E,2007ApJ...656..653L,2008ApJ...676...10E,2016AA...594A..10P} assume instead a functional form for the emission of each astrophysical emission, and fit the value of its parameters in a global optimization approach. However, the performance of such ``non-blind'' methods highly depends on the reliability of any prior information, in particular on the functional forms of the frequency dependence of foreground emission, and possibly on priors on their parameters. 

As an alternative, several ``blind'' approaches have been developed. Those only assume a structure for the component mixture (e.g. linear superposition), and typically rely on the statistical independence of components of various physical origins for their separation in multi-frequency observations. Such blind methods include Independent Component Analysis (ICA) methods, based on the maximization of some measure of the non-Gaussianity of independent sources \citep{2004MNRAS.354...55B}; the Spectral Matching Independent Component Analysis (SMICA), which uses spectral diversity and decorrelation to identify independent components~\citep{2003MNRAS.346.1089D,2005EJASP2005..100M,2005MNRAS.364.1185P,2008arXiv0803.1814C,2009A&A...503..691B};
and the (conceptually close) Correlated Component Analysis (CCA) method \citep{2006MNRAS.373..271B}. 

Variants of yet another component separation method, the so-called Internal Linear Combination method (ILC), have been extensively used for the analysis of WMAP and Planck survey data to obtain foreground-cleaned CMB maps and power spectra~\citep{1996MNRAS.281.1297T,2003PhRvD..68l3523T,2003ApJS..148...97B,2004ApJ...612..633E,2006ApJ...645L..89S,2009A&A...493..835D,2012MNRAS.419.1163B,2013MNRAS.435...18B}. The method can be adapted to also extract maps of other astrophysical components \citep{2011MNRAS.410.2481R,2011MNRAS.418..467R}. Closely related approaches based on foreground template matching, including the Spectral Estimation Via Expectation Maximisation (SEVEM)~\citep{2003MNRAS.345.1101M,2012MNRAS.420.2162F}, have been used in the analysis of data from previous CMB surveys. Finally, methods based on sparsity rely on the representation of foreground emission with a limited number of coefficients in a redundant basis of functions~\citep{2007ITIP...16.2662B}. The component separation problem in the context of CMB observations and the comparison of performances of various approaches have been discussed in several dedicated papers~\citep{2008A&A...491..597L,2009LNP...665..159D,dunkley2009}.

\subsection{The ABS method}
\label{sec:abs}
A method to directly estimate the angular power spectrum of the CMB from the multivariate angular power spectrum of multi-frequency observations was proposed by~\cite{2016arXiv160803707Z}. This method, called Analytical method of Blind Separation (ABS), shares conceptual elements with SMICA (both methods estimate the CMB angular power spectrum first, before any separation of component maps), as well as with the ILC 
(see Sec.~\ref{sec:ilc}) 
and with GNILC (see Sec.~\ref{sec:gnilc}: both methods estimate a subspace of significant components). Advantages of the ABS method include a simple implementation which does not involve heavy computations, and the fact that it does not rely on any assumption about the characteristics of the foreground components. It uses the measured cross-band power between different frequency bands, from which the CMB power spectrum can be solved analytically, avoiding multiple parameter fitting procedure. A description of the mathematical formalism and numerical techniques was provided by~\cite{2016arXiv160803707Z}. The ABS method was also tested against simulated temperature maps observed by the Planck space mission ~\citep{yao2018}. As an extension of this work, we test here the ABS method on simulated CMB polarization observations with a sensitive future space mission, targeting the reconstruction of the $E$- and $B$-mode power spectra. We also clarify it's connection to other classical component separation methods, and discuss the origin of possible biases in the power spectrum estimation.

The ABS formalism assumes a data model in which the observations in $N_f$ different frequency channels, $\vec d^{\rm obs}(\ell,m)$, contain a superposition of CMB, foreground emission, and noise as follows: 
\begin{equation}
    \vec d^{\rm obs}(\ell,m) = \vec f d^{\rm cmb}(\ell,m) + \vec d^{\rm fore}(\ell,m) + \vec d^{\rm noise}(\ell,m)\,,
\end{equation}
where $\vec f =[f_1,...,f_{N_f}]^T$ is the ``mixing column'' (or ``mixing vector'') of the CMB, and $d^{\rm cmb}$, $\vec d^{\rm fore}$, and $\vec d^{\rm noise}$ refer to CMB, total foreground emission, and noise respectively. Such a data model can be written independently for each of the $T$, $E$ and $B$ fields.
Without loss of generality, we can use thermodynamic units for the observations, so that the CMB emission pattern is constant across frequencies, and $f_i=1$,  $\forall i$. Alternatively, we can also use units in which the noise has unit variance, i.e., re-scale the observations by dividing the data $ d_i^{\rm obs}(\ell,m)$ from each channel $i$ by its noise level $\sigma_i^{\rm noise}(\ell)$ (assumed here to be isotropic, and hence independent of $m$, for simplicity).

Computing the empirical multivariate power spectrum of the observations (in whatever units), we get 
\beq
\mathcal{D}^{\rm obs}_{ij}(\ell) = f_if_j\mathcal{D}^{\rm cmb}(\ell) + \mathcal{D}^{\rm fore}_{ij}(\ell) + \mathcal{D}_{ij}^{\rm noise}(\ell)\,,
\eeq
where the data, and hence also the vector $\vec f =[f_1,...,f_{N_f}]^T$, are expressed in the most convenient units (i.e., thermodynamic units, or noise units for noisy observations). $\mathcal{D}^{\rm obs}_{ij}(\ell)$ represents the cross-band power spectrum of the (polarized) observations in the $i$- and $j$-th frequency channels. The three main contributions to the data are the CMB signal, $\mathcal{D}^{\rm cmb}$; the auto and cross band power matrix of the foregrounds, $\mathcal{D}_{ij}^{\rm fore}$; and the noise contribution to the auto and cross-spectra, $\mathcal{D}_{ij}^{\rm noise}$. Note that this assumes vanishing correlation between any two of the maps of CMB, foreground emissions, and noise in any of the channels. While this is a safe assumption from an ensemble average prospective, in view of the very different physical origins of those contributions to the total signal, for a single finite data set empirical correlations exist and, as we will see later, impact the performance of a method such as ABS, which relies on the decorrelation of the main components in the observations.

In a first step, we subtract from this multivariate spectrum an estimate of the noise contribution (the ensemble-average of the instrumental noise spectrum is assumed to be known). We get:
\beq\label{eq:data}
\mathcal{D}^{\rm obs}_{ij}(\ell) = f_if_j\mathcal{D}^{\rm cmb}(\ell) + \mathcal{D}^{\rm fore}_{ij}(\ell) + \delta \mathcal{D}_{ij}^{\rm noise}(\ell)\,,
\eeq
where now $\delta \mathcal{D}_{ij}^{\rm noise}(\ell)$ is the residual of instrumental noise contribution after noise de-biasing, and is centred around zero. Assuming an uncorrelated instrumental Gaussian noise distribution, with zero mean and rms level $\sigma_{\mathcal{D},i}^{\rm noise}$ for the $i$-th frequency channel, the properties of the residual noise are:
\begin{eqnarray}\label{eq:noise}
&&\left<\delta \mathcal{D}_{ij}^{\rm noise}\right> =0\,,\nonumber \\
&&\left<(\delta \mathcal{D}_{ij}^{\rm noise})^2\right> = \frac{1}{2} \sigma_{i}^{\rm noise} \sigma_{j}^{\rm noise} (1+\delta_{ij})\,,
\end{eqnarray}
where $\delta_{ij}$ is the Kronecker symbol, and $\forall i$ we have $\sigma_{i}^{\rm noise}=1$ when we work with noise-whitened observations\footnote{Note that this amounts to changing the units of the input maps, and hence also the CMB mixing column \vec f.} (which we do in the present paper).
The multivariate power spectrum, for each $\ell$, is a semi-positive symmetric matrix. It can be diagonalized as
\beq\label{eq:diagonalization}
\mathcal{D}^{\rm obs}_{ij}(\ell) = {\bf {E}} \Lambda {\bf {E}}^T,
\eeq
where the columns of $\bf {E}$ are an orthonormal basis of eigenvectors, and $\Lambda = {\rm diag}({\lambda_\mu})$ is a diagonal matrix. 

\cite{2016arXiv160803707Z} show that the CMB variance for each $\ell$ is obtained as
\beq\label{eq:abs}
\widehat{\mathcal{D}}^{\rm cmb} = \left( \sum_{\mu=1}^{M+1} G^2_{\mu}\lambda_{\mu}^{-1}\right)^{-1}\,,
\eeq
where $\vec G = {\bf {E}^T} \vec f$, i.e., $G_\mu = \sum_i E_{i \mu} f_i$, and $M$ is the rank of $\mathcal{D}_{ij}^{\rm fore}$, or the rank of the subspace spanned by foreground components (i.e. the number of component maps that are required to represent the total foreground emission in any of the observation frequency channels as a linear mixture of those components). The sum in Eq.~\ref{eq:abs} extends from 1 to $M + 1$ to account for $M$ dimensions of foreground emission and one extra dimension for CMB (which is not in the foreground subspace when the observations comprise enough frequency channels, in frequency bands that break any possible degeneracy). 
The ABS method, however, thresholds the eigenvalues $\lambda_\mu$, to keep only those that are not dominated by noise variance. 
After thresholding, the power spectrum estimate is implemented in practice as
\beq\label{eq:abs2}
\widehat{\mathcal{D}}^{\rm cmb} = \left( \sum^{{\lambda}_{\mu}\geq \lambda_{\rm cut}} G^2_{\mu}\lambda_{\mu}^{-1}\right)^{-1}\,,
\eeq
which amounts to keeping only in the sum those eigenvalues which are not dominated by noise, i.e., those that contain significant foreground or CMB signal. Hence, Eq.~\ref{eq:abs2} gives a solution that is almost equivalent to Eq.~\ref{eq:abs}.

\cite{yao2018} have shown that the estimate of the CMB power spectrum is not very sensitive to the value of the threshold in the range $0.5 \lesssim  {\lambda}_{\rm cut} \lesssim  1$ (when the system is written for noise-whitened data). In the present paper, we adopt a threshold ${\lambda}_{\rm cut} = 0.5$.

As a final point, we note that the thresholding step introduces an additional technicality. While thresholding allows one to  
discard noise-dominated eigenmodes, it can also discard part of the signal of interest, and hence introduce a bias. This is the case, in particular, when one works in a low signal-to-noise regime, as is often the case for the observation of the faint CMB $B$-modes. To avoid this potential issue, the ABS method proposed by \cite{2016arXiv160803707Z} introduces a ``shift parameter'' $\mathcal{S}$. Before diagonalization and thresholding, the covariance of the observations is ``shifted'' to
\beq
\widetilde{\mathcal {D}}^{\rm obs}_{ij}(\ell) = \mathcal{D}^{\rm obs}_{ij}(\ell) + f_i f_j \mathcal{S}.
\label{eq:absshift}
\eeq
This corresponds to adding to the covariance of the observations an extra CMB term with power $\mathcal{S}$ in CMB units. For large $\mathcal{S}$, this guarantees that the CMB signal is strong as compared to the noise, and thus ``lives'' in the subspace spanned by the eigenvectors that are not cut-out in the thresholding process, i.e., the set of eigenvectors that are kept in the summation used for estimating the CMB power (Eq.~\ref{eq:abs2}). The amount of CMB power that was artificially added to the data is then subtracted from the final ABS power spectrum estimate in the last step of the analysis.

\subsection{Connection of ABS to ILC}
\label{sec:ilc}
The ILC method estimates a clean CMB component from multi-frequency observations as the linear combination
\beq
\widehat{d}^{\rm cmb} = \vec w^T {\vec d}^{\rm obs} = \frac{\vec f^T [{\bm{\mathcal D}}^{\rm obs}]^{-1}}
{\vec f^T [{\bm{\mathcal D}}^{\rm obs}]^{-1}\vec f} \, {\vec d}^{\rm obs}.
\eeq
For each $\ell$, the variance of the ILC solution is simply the power spectrum of the ILC map for that harmonic mode, which can be used as an estimator for the CMB power spectrum. We get
\beq
\widehat{\mathcal D}^{\rm cmb} = \vec w^T {\bm{\mathcal D}}^{\rm obs} \vec w 
= \frac{1}{{\vec f^T [{\bm{\mathcal D}}^{\rm obs}]^{-1}\vec f}}.
\eeq
Using Eq.~\ref{eq:diagonalization}, and noting that $[{\bm{\mathcal D}}^{\rm obs}]^{-1} = {\bf {E}} \Lambda^{-1} {\bf {E}}^T$, we get
\beq\label{eq:PS-ILC}
\widehat{\mathcal D}^{\rm cmb} = \frac{1}{{\vec f^T {\bf {E}} {\bm{\Lambda}}^{-1} {\bf {E}}^T \vec f}}
= \left( \sum_{\mu=1}^{N_f} G^2_{\mu}\lambda_{\mu}^{-1}\right)^{-1} \,,
\eeq
which is strongly reminiscent of Eqs.~\ref{eq:abs} and~\ref{eq:abs2}  with, however, two differences: 
the first is in the range of eigenvectors and eigenvalues which are kept in the summation, as the ILC keeps all $N_f$ dimensions, while ABS keeps only the most significant ones (with signal-to-noise ratio above some threshold); the second is that, in Eq.~\ref{eq:PS-ILC}, the eigenvalues $\lambda_{\mu}$ have not been de-biased from the noise. Hence, the power spectrum that is computed directly from the ILC map must be corrected by subtracting the projected noise. Simple algebra shows that this amounts to subtracting 1 to each eigenvalue on the diagonal of the ${\bm{\Lambda}}$ matrix. We hence recover, after noise de-biasing, the exact ABS solution in the case where $M+1 = N_f$ (i.e., when there is significant signal in all dimensions of the diagonalized system).

\subsection{Connection of ABS to GNILC}
\label{sec:gnilc}

We note that the diagonalization and thresholding that are performed in the implementation of ABS are very similar to those that are performed in the GNILC method of \cite{2011MNRAS.418..467R}, with the exception that the latter considers that the ``noise'' term can contain also astrophysical components of known multivariate spectrum. In that case, the implementation of GNILC in harmonic space (instead of needlet space) would be an extension of the harmonic-space ILC which could be used to separate different astrophysical foreground emissions in harmonic domain, and share many common features with the ABS method.

\section{Biases in the ABS power spectrum estimation}
\label{sec:biases}

For any set of observations $\vec d^{obs}(\ell,m)$, the data in each channel can be expressed as a linear mixture of at most $N_f$ different maps, as follows:
\begin{equation}
    \vec d^{\rm obs}(\ell,m) = {\bf {A}} \vec d^{\rm comp}(\ell,m),
\end{equation}
where $\vec d^{\rm comp}(\ell,m)$ is the vector of components present in the observations. Note that such a decomposition is \emph{always} possible, at least with ${\bf A}$ equal to the identity matrix, and $\vec d^{\rm comp}(\ell,m) = \vec d^{\rm obs}(\ell,m)$. This trivial decomposition, however, is of little practical interest. The key ideas behind ABS (and several component separation methods) are that there are physical components which are superimposed in each frequency map; that the CMB is uncorrelated with the other components (foregrounds, and noise if any); and that the scaling of the CMB with frequency is known and universal, i.e. does not depend on the direction on the sky (rigid scaling: the CMB map observed at one frequency is a simple scaling of the CMB map observed at any other frequency, when both observations are at the same angular resolution).

Consider the $N_f$-dimensional space of possible frequency scaling laws for the available observations. We can single-out the direction given by the ``mixing vector'' of the CMB (i.e. the vector $ \vec f=[1,1, ...,1]^t$ when the observations are expressed in CMB units). We can also decompose the total foreground emission $\vec d^{\rm fore}(\ell,m)$ as a linear mixture of a set of independent component maps with respective scaling vectors $\vec a_i$, $i=\{1,..,M\}$.

In the ideal application case of ABS, the observed frequency maps are noiseless, the dimension $M$ of the foreground space is such that $M+1 \leq N_f$, the CMB direction (or mixing vector) is not in the $M$-dimensional foreground subspace, and the correlation of CMB with foregrounds and noise exactly vanishes. When this is the case, it is possible, in theory, to perfectly separate the CMB signal from foreground signals, and hence get an unbiased CMB power spectrum estimate with ABS. In real observations of a single sky, complexities arise.

\subsection{Residual contamination by foreground emission}

In practice, foreground emission is a superposition of many distinct processes with different emission laws. Each region of galactic emission, each extragalactic source, contributes to the observations with its own emission law. Hence, for real observations, the foreground subspace is that of the original data, and $\mathcal{D}^{\rm fore}_{ij}(\ell)$, has rank $N_f$. The CMB being in the space spanned by foreground components, it is impossible to get a CMB spectrum which would not contain at least a bit of positive bias from foreground emission power along the CMB direction (i.e. mixing vector). 

This contamination is minimised by increasing the number of frequency channels (and optimising their distribution in frequency). %
Most relevant astrophysical emission laws are smooth functions of frequency.\footnote{For completeness, we note that emission from a few atomic or molecular lines, such as CO emission, must also be taken into account~\citep{2013A&A...553A..96D,2014A&A...571A..13P}; such emissions, however, impact only a subset of the frequency channels used in the observations.} 
If these functions are sampled with a sufficient number of frequency bands, it is possible to \emph{approximately} predict the emission in the frequency channel(s) most sensitive to the CMB using a linear combination of observations at other frequencies, which serve as foreground monitors. When the error in this ``prediction'' is smaller than the noise, one can subtract the foreground contamination down to the noise level (and thus get a reasonably good foreground-cleaned CMB map). 

In the ABS formalism, when this is the case, only a subset of the observations is sufficient to model the foregrounds at any frequency. The rank of the covariance matrix of \emph{significant} foregrounds is $M<N_f$ (other eigenvalues of $\mathcal{D}^{\rm fore}$ being smaller than the noise variance).
A good choice of frequency channels is when the mixing vector of the CMB is not in the ``foregrounds subspace'' spanned by the eigenvectors of the foreground covariance matrix $\mathcal{D}^{\rm fore}$ corresponding to large eigenvalues of foreground power.

Residuals of foreground contamination \emph{bias high} the CMB power spectrum estimate with ABS. With enough frequency channels and enough sensitivity, this bias can be kept within acceptable values. However, the exact requirements are difficult to evaluate without an accurate model of foreground emission.


\subsection{The finite sample bias}

The covariance matrix of the observations $\mathcal{D}^{\rm obs}$ is computed using a finite sample of observations. For this reason, even if the CMB is theoretically not correlated with noise and foregrounds, empirical correlations (and anti-correlations) exist. As discussed at length in the appendix of \cite{2009A&A...493..835D}, these empirical correlations result in a loss of modes in the reconstructed ILC CMB map. The ABS power spectrum estimate, being similar to the noise-debiased power spectrum of the ILC map (and if all the eigenvalues of the covariance matrix of the observations are above the threshold, equal to it), suffers from the same bias. This bias can be reduced by increasing the number of modes that are used to compute the covariances, for instance by binning in $\ell$ space.

Empirical correlations between finite-size samples \emph{bias low} the CMB power spectrum estimate with ABS.

\subsection{Thresholding biases}

The instrumental noise can lead to noise-dominated eigenmodes, with eigenvalues $|\lambda_\mu| \lesssim 1$ in $\mathcal{D}^{\rm obs}_{ij}$. Rejecting positive eigenvalues in the summation of Eq.~\ref{eq:abs2} reduces the value of the sum, and hence increases the estimated power spectrum, while rejecting negative eigenvalues does the opposite. 

In the case where there are $M<N_f$ foreground components and one CMB component, all of which associated to eigenvalues that are large as compared to those of the noise, one can select the threshold $\lambda_{\rm cut}$ such that all the noise-dominated modes are discarded, and contributions from negative and positive rejected eigenvalues cancel out on average. One must avoid discarding an excess of negative or positive eigenvalues to avoid a thresholding bias on the CMB power spectrum estimate with ABS. 

\subsection{The choice of a shift parameter}

The free-shift parameter $\mathcal{S}$ shifts the amplitude of the input CMB power spectrum from $\mathcal{D}^{\rm cmb}(\ell)$ to $\mathcal{D}^{\rm cmb}(\ell) + \mathcal{S}$. This changes the eigendecomposition of $\mathcal{D}^{\rm obs}$. In particular, it helps keeping the CMB in the subspace that is not rejected by thresholding.
In our calculations, we use $\mathcal{S} = 1000\sigma_{\mathcal{D}}^{\rm noise}$ as a baseline, unless specified otherwise. In the case with polarized AME, discussed in Appendix~\ref{app:AME}, we use $\mathcal{S} = 100\sigma_{\mathcal{D}}^{\rm noise}$. 


\section{Simulations and analysis pipeline}
\label{sect:map}

\subsection{Simulated data sets}

\begin{figure}[!htpb]
\centering
\includegraphics[width=3.2in]{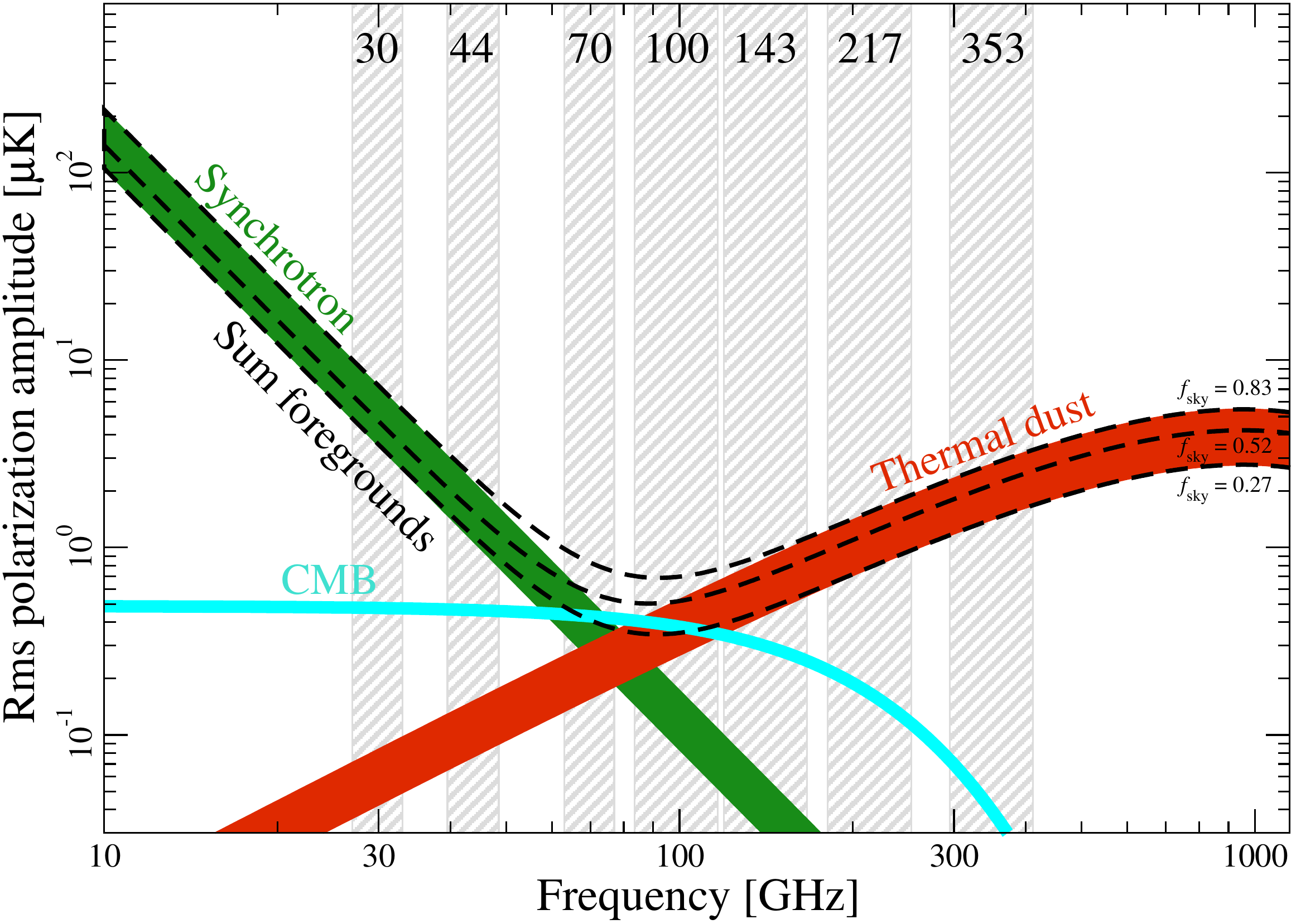}
\caption{Frequency dependence and relative levels of the polarized foregrounds and CMB signal (dominated by polarization E-modes). The Planck experiment frequency channels are explicitly shown in the grey shaded region (https://www.cosmos.esa.int/web/planck/picture-gallery).} 
\label{planck_pol} 
\end{figure}

We test the ABS method on simulated data sets which could be obtained with a future sensitive CMB polarization space mission.
We consider an experiment with ten frequency bands, the main characteristics of which are listed in Table~\ref{tab:exp}. The experiment frequency range covers the maximum of CMB emission relative to astrophysical foregrounds, with extra channels at low and high frequency to monitor contamination by synchrotron emission, which dominates the low-frequency part of the spectrum, and by thermal dust, which dominates the higher part (see Figure~\ref{planck_pol}). The 10 frequency channels are a subset of those in the experimental setup developed for the Probe of Inflation and Cosmic Origins (PICO) experiment\footnote{https://zzz.physics.umn.edu/ipsig/start}.

Theoretical CMB power spectra ($E$- and $B$-modes) are obtained with the publicly available CAMB software~\citep{Lewis2000}, with parameters from the cosmological model fitting best Planck observations~\citep{2016AA...594A..13P}. Polarized CMB $Q$ and $U$ maps are generated using the LensPix software~\citep{lenspix, lenspix1}. We consider two limiting cases for primordial tensor modes, $r=0$ and $r=0.05$. For each frequency channel, we add to the mock observations emission from two polarized foreground components (thermal dust and synchrotron), generated with the \texttt{PySM} model~\citep{2017MNRAS.469.2821T}. We use the default \emph{d1s1} model, in which the dust emission law is modelled with a single modified blackbody with spectral index and temperature varying across the sky, and the synchrotron emission law in each sky pixel is modelled as a power law with, again, spectral index varying across the sky.
Gaussian instrumental noise, uncorrelated from pixel to pixel and from channel to channel, at the level indicated in Table~\ref{tab:exp}, is also included in the simulations. For all of this work, we use the HEALPix~\citep{2005ApJ...622..759G} pixelization scheme at ${N_{\rm side}= 1024}$. 

\begin{figure*}[htpb]
\centering
\mbox{

 \subfigure{
   \includegraphics[width=2.3in] {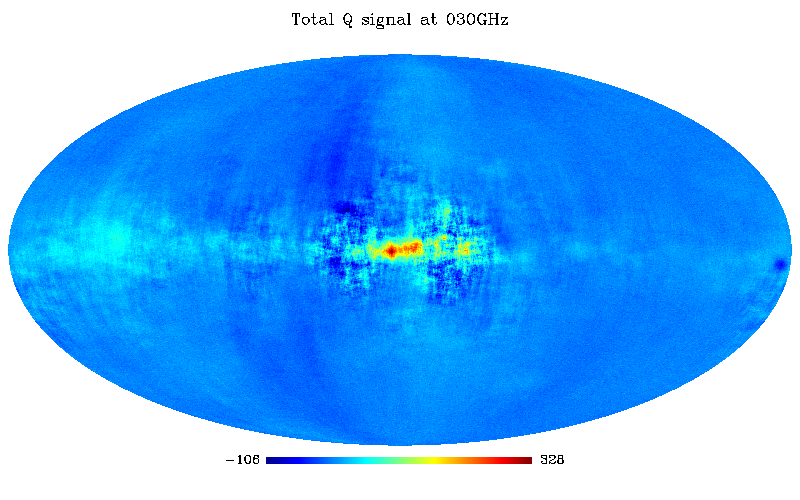}
   }

 \subfigure{
   \includegraphics[width=2.3in] {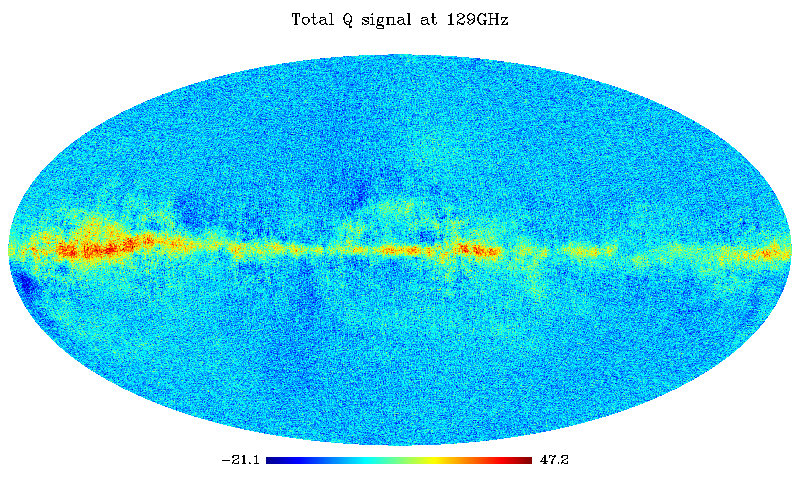}
   }

 \subfigure{
   \includegraphics[width=2.3in] {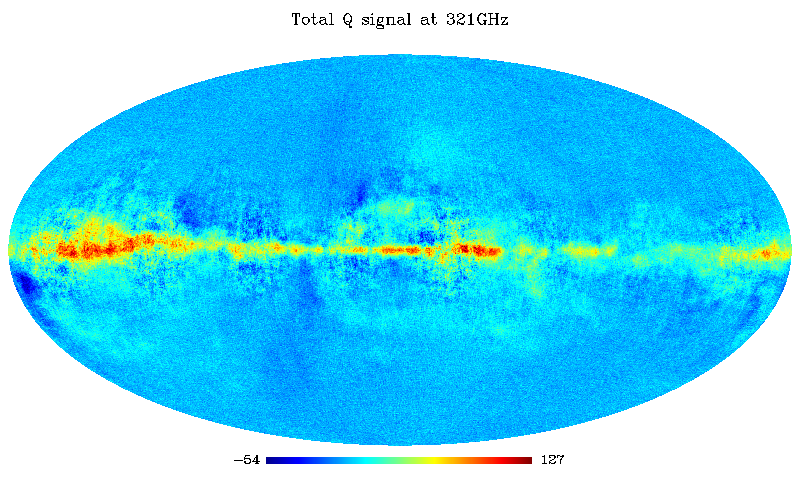}
   } } \\
   
  \mbox{
 \subfigure{
   \includegraphics[width=2.3in] {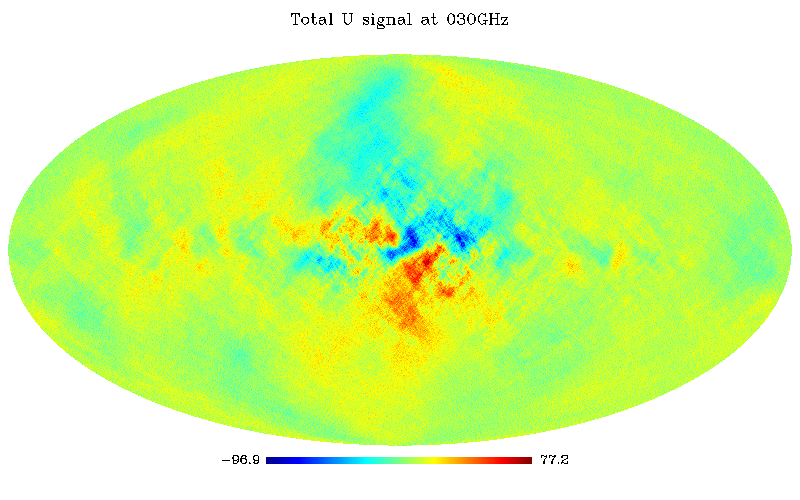}
   }

 \subfigure{
   \includegraphics[width=2.3in] {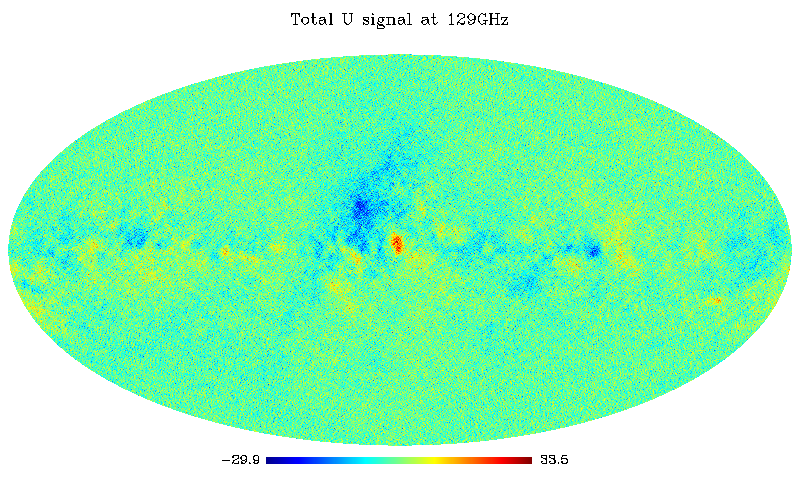}
   }

 \subfigure{
   \includegraphics[width=2.3in] {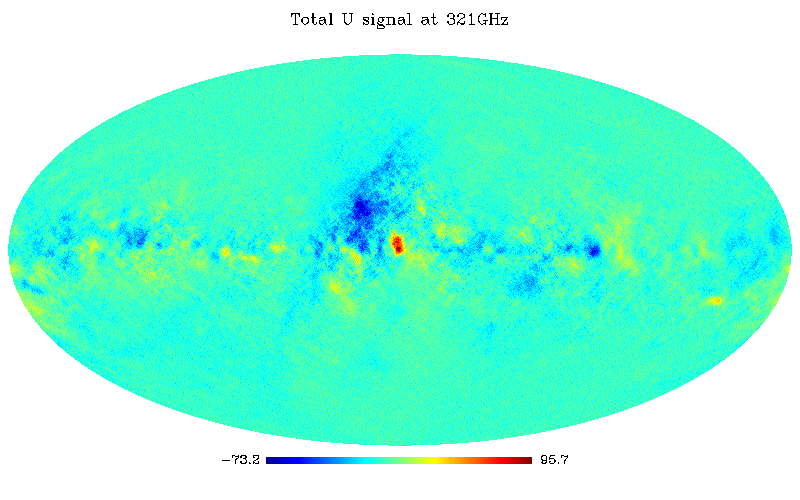}
   }

}

\caption{The  total $Q$ (upper panel) and $U$ (lower panel) simulated full-sky maps (without any beam convolution): CMB + synchrotron + thermal dust + noise. The dust and synchrotron are from the PySM model \emph{d1s1} (see text). From left to right: 30, 129, and 321 GHz. The scale is in $\mu$K$_{\rm CMB}$.}
\label{fig_maps}
\end{figure*}

\begin{figure*}[htpb]
\centering
\mbox{

 \subfigure{
   \includegraphics[width=2.3in] {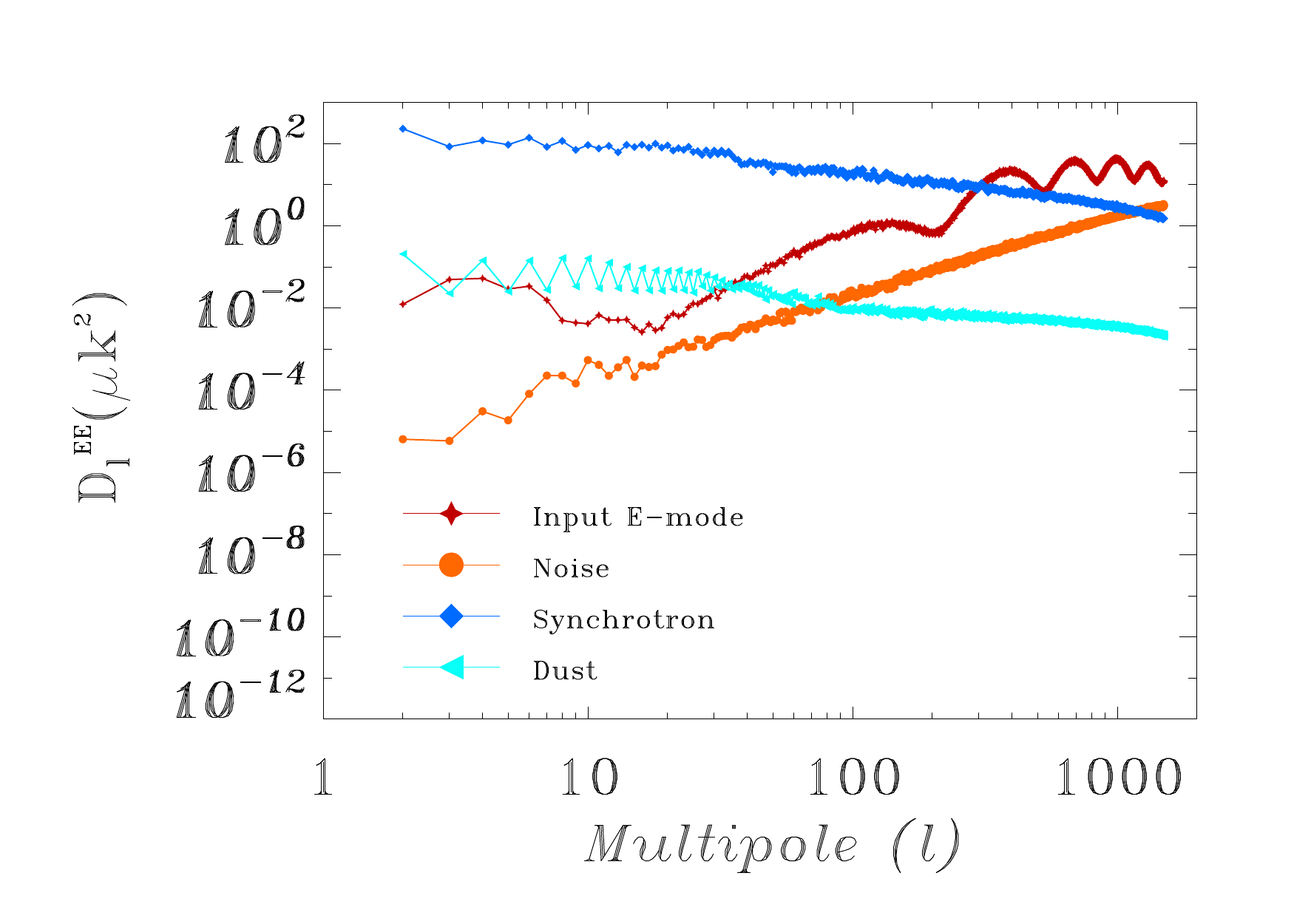}
   }

 \subfigure{
   \includegraphics[width=2.3in] {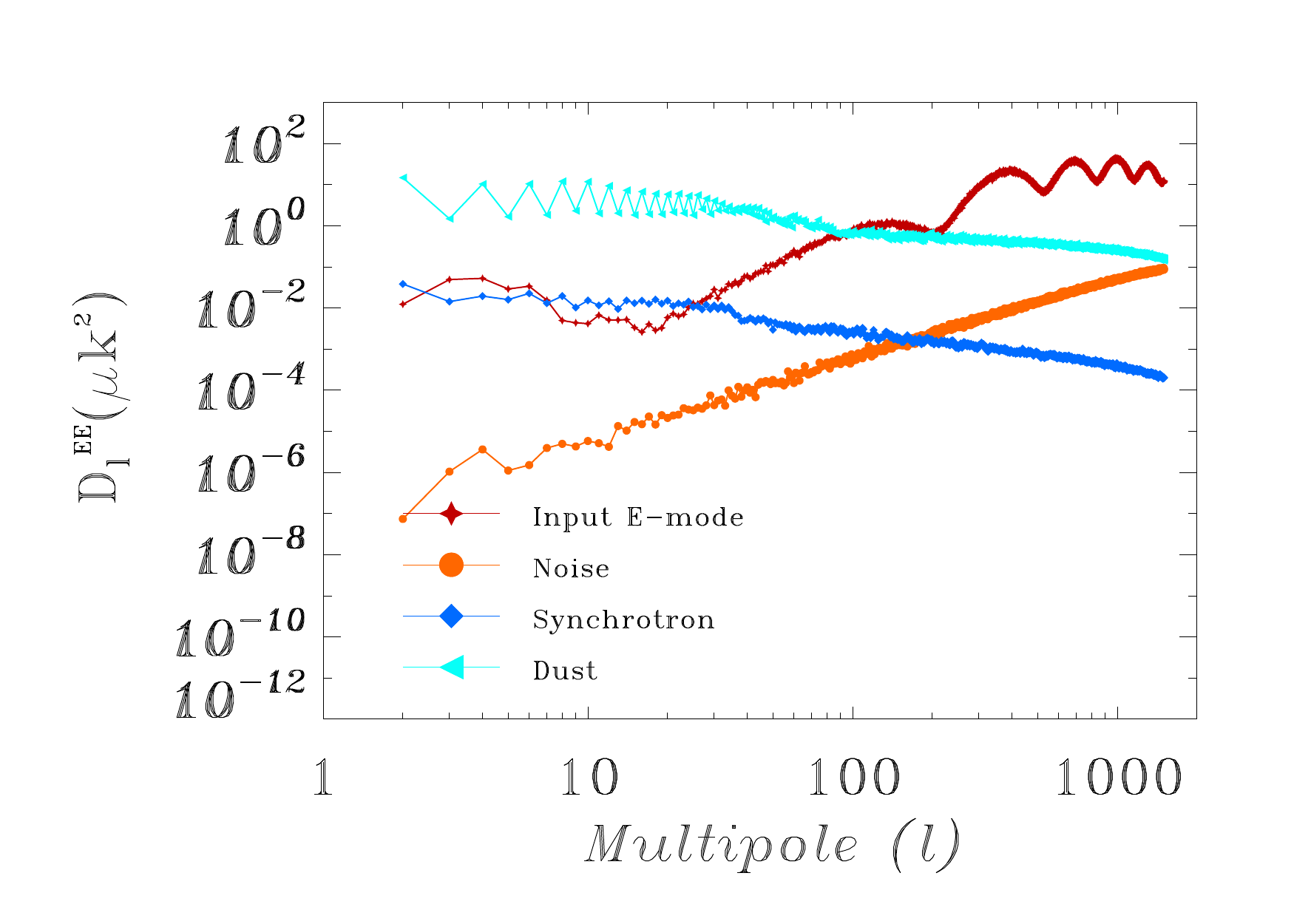}
   }

 \subfigure{
   \includegraphics[width=2.3in] {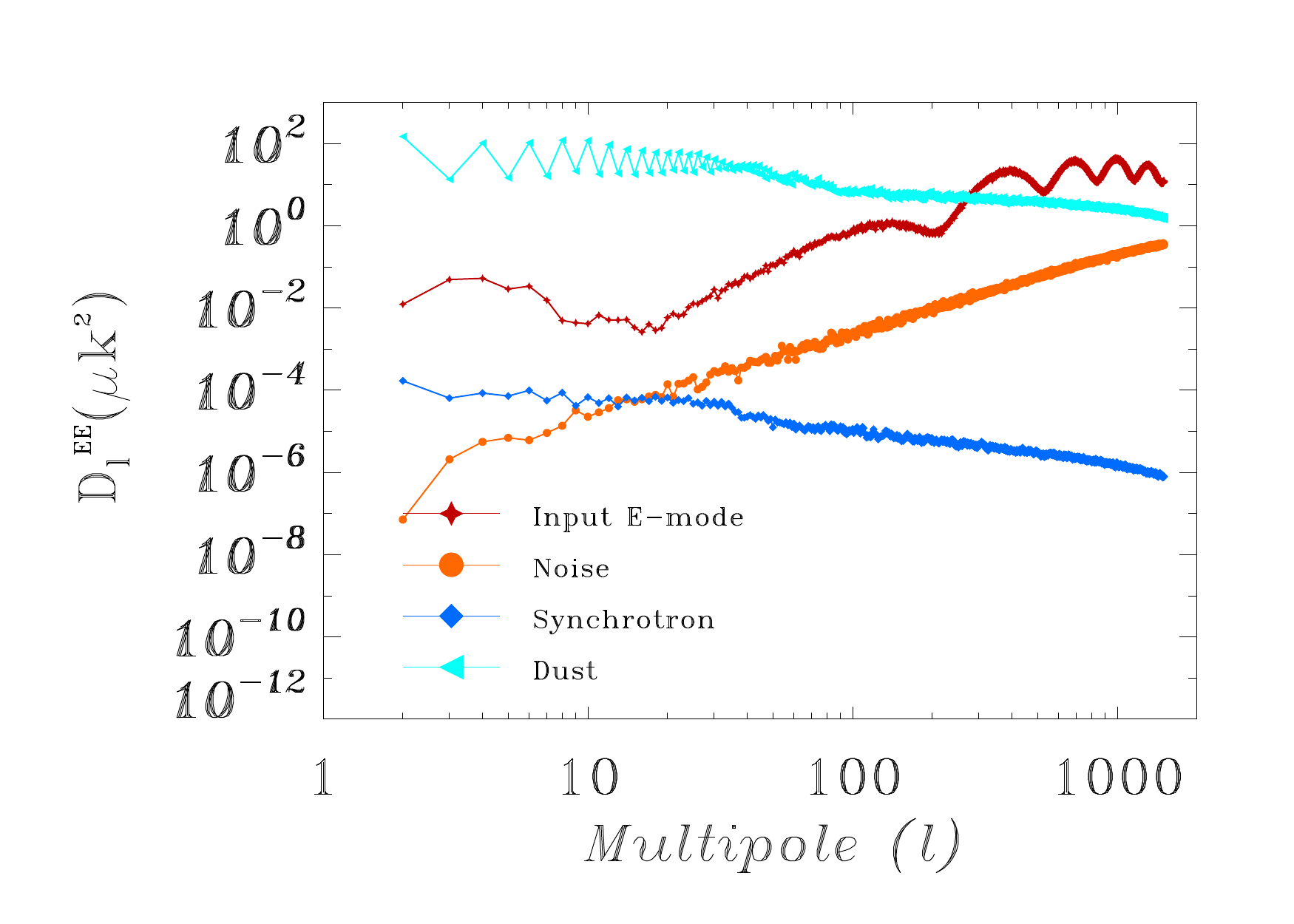}
   } } \\
   
  \mbox{
 \subfigure{
   \includegraphics[width=2.3in] {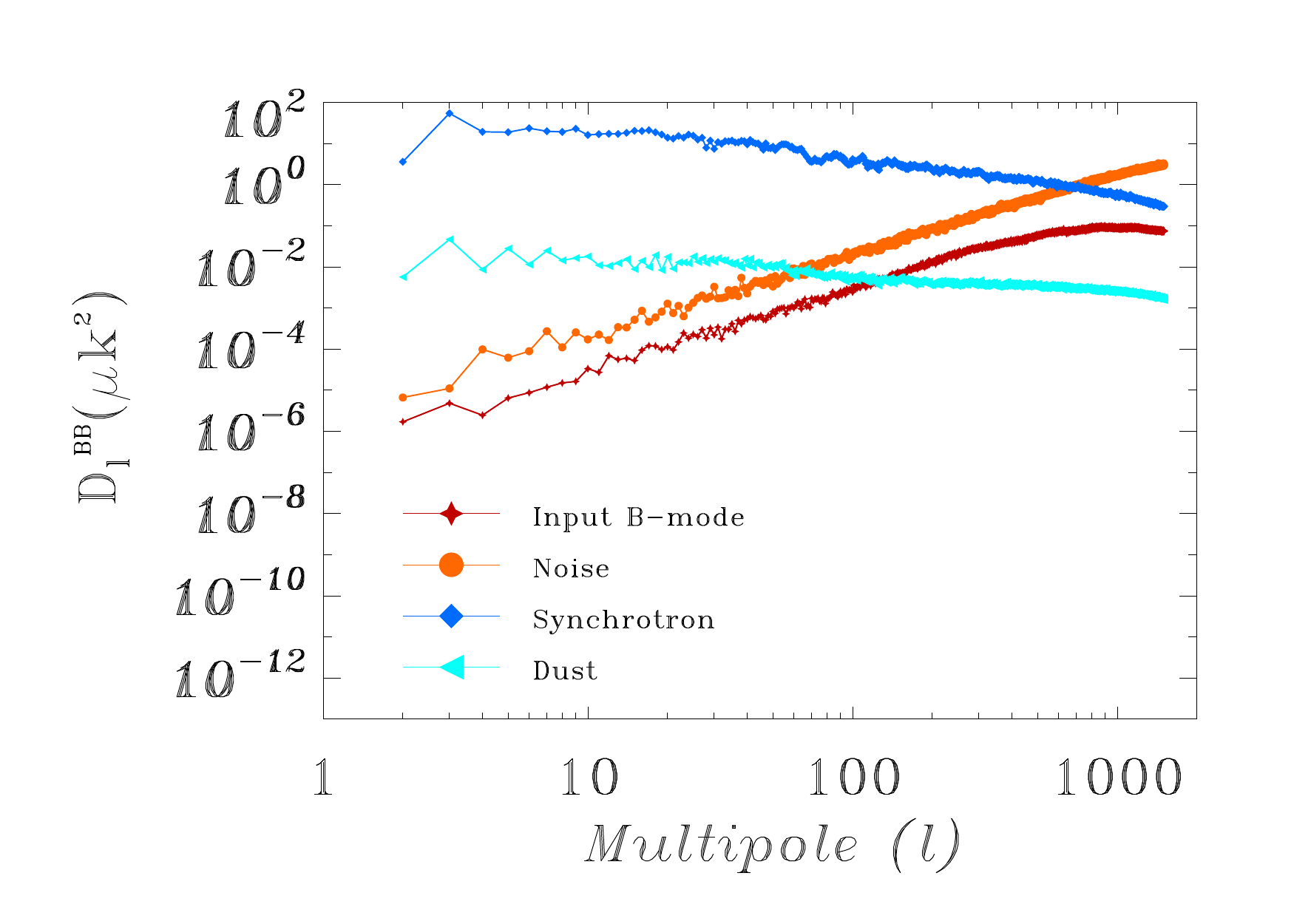}
   }

 \subfigure{
   \includegraphics[width=2.3in] {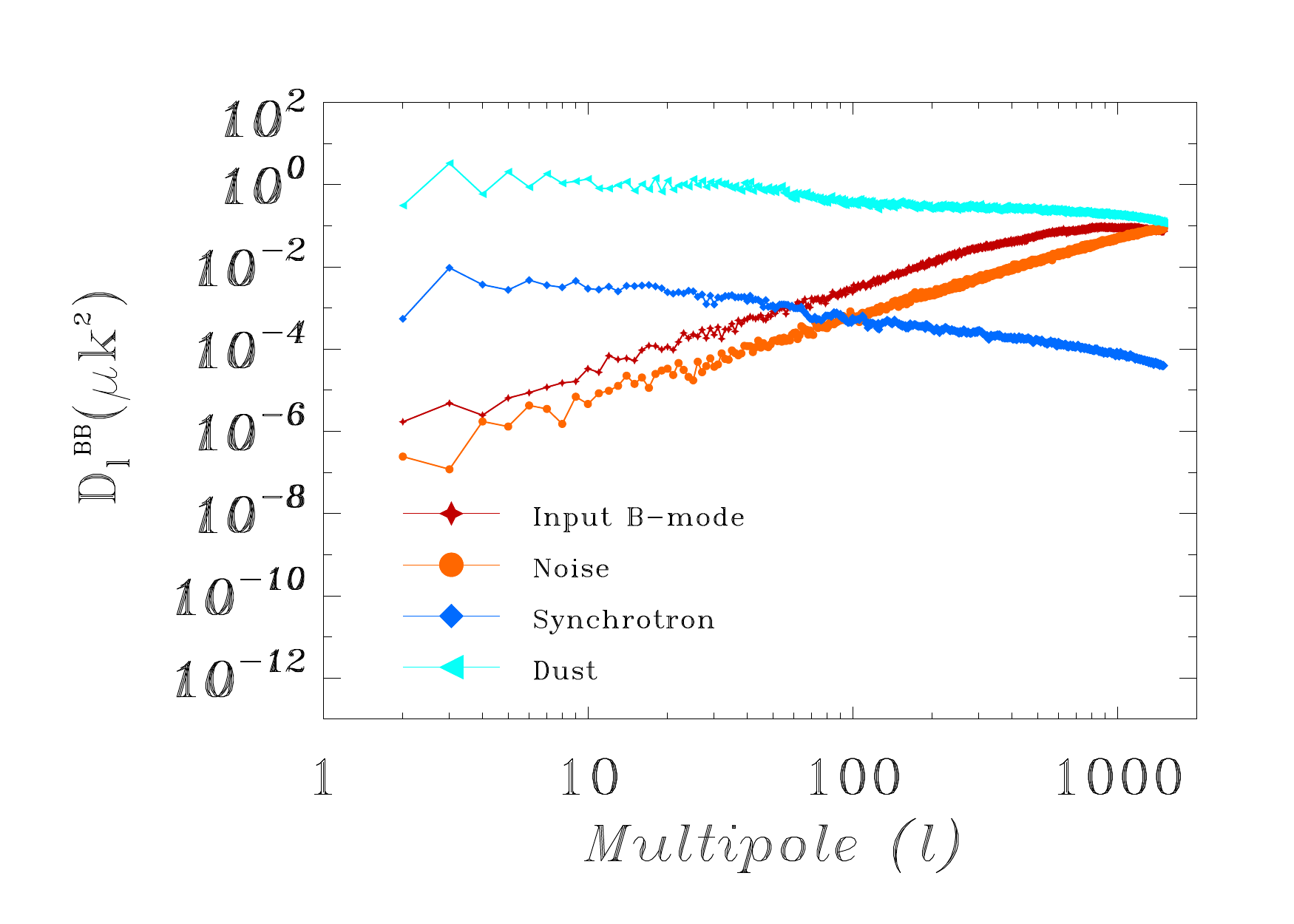}
   }

 \subfigure{
   \includegraphics[width=2.3in] {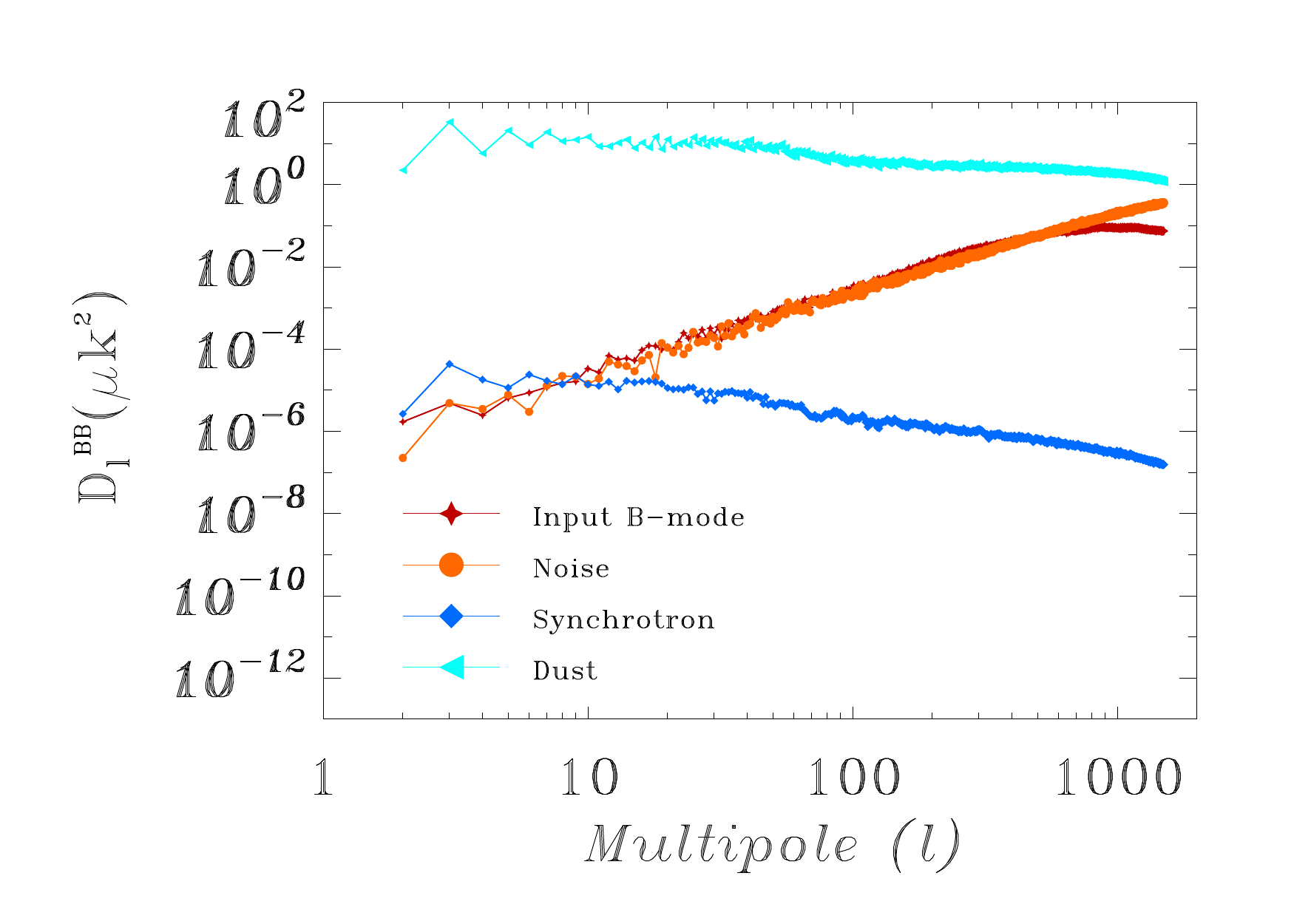}
   }

}

\caption{The $E$- and $B$-mode angular power spectra of the simulated full-sky maps (without any beam convolution), with a comparison of all the simulated components for three frequency channels: CMB, noise, synchrotron and thermal dust. From left to right: 30GHz, 129GHz, and 321GHz.}
\label{fig_PS}
\end{figure*}

Polarization maps for the $Q$ and $U$ Stokes parameters, as well as $E$- and $B$-mode power spectra, for three representative frequency bands, without any beam convolution, are displayed in Figures \ref{fig_maps} and \ref{fig_PS}.

\begin{table}
\begin{center}
\caption{Experimental setup considered in this study.}
\label{tab:exp}
\begin{tabular}{ccc}
\hline
Band center & Beam FWHM&  noise level\\
(GHz)     &   (arcmin)       & ($\mu$K${}_{\rm CMB}$-arcmin) \\
\hline
030 &  28.3  & 12.4 \\
043   & 22.2  & 7.9 \\
075   & 10.7   & 4.2  \\
090   &  9.5  & 2.8  \\
108   &  7.9  & 2.3  \\
129  &  7.4  & 2.1  \\
155  &  6.2  & 1.8 \\
223  &  3.6  & 4.5  \\
268  &  3.2  & 3.1  \\
321  &  2.6  & 4.2  \\
\hline
\end{tabular}
\end{center}
\end{table}

\subsection{CMB $E$/$B$ decomposition}
\label{CMB $B$-mode decomposition}

In the ideal case of full-sky coverage, the CMB linear polarization field, described by the Stokes parameters $Q(\hat n)$ and $U(\hat n)$, can be unambiguously decomposed into $E$-mode and $B$-mode components \citep{zaldarriaga-b-mode, zal+seljak:1997, kamionkowski97,kamionkowski-b-mode}.
 However, if the polarization field is not measured over the full sky, the decomposition into $E$ and $B$ modes is not unique, since there are modes that satisfy the properties of both $E$ and $B$ modes simultaneously over the observed sky patch \citep{2001PhRvD..64f3001T}. This generates confusion between $E$ and $B$ modes, which must be corrected for to correctly estimate the $E$- and $B$-mode power spectra~\citep{ebmixture0,ebmixture2,ebmixture3,ebmixture4,ebmixture5,ebmixture6,ebmixture9,ebmixture10,ebmixture11,ebmixture12,liu2018, 2liu2018, liu2019,2liu2019,3liu2019}, and to carefully take into account residuals that may still be present in the maps~\citep{larissa2017}.

\begin{figure}[!htpb]
\centering
\includegraphics[width=3.2in] {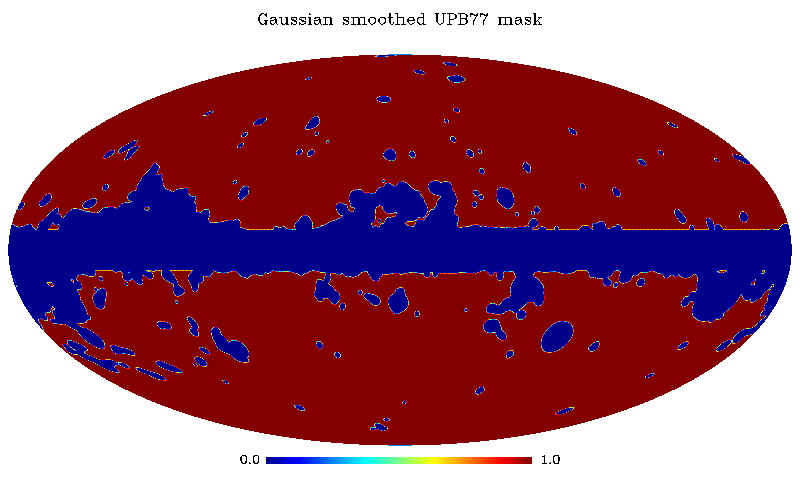}
\caption{The apodized UPB77 mask used for partial-sky analysis in this work.} 
\label{mask_planck} 
\end{figure}

For this work, part-sky analyses are made using the UPB77 Planck mask, smoothed with a Gaussian smoothing kernel that extends to a distance $\delta_c$. Specifically, the smoothing function is
\begin{equation}\label{gauss}
W(\delta)=\left \{
\begin{aligned}
&
\frac{1}{2}+\frac{1}{2}{\rm erf}\bigg(\frac{\delta-\frac{\delta_c}{2}}{\sqrt{2}\sigma }\bigg)  &\delta<\delta_{c}
\\
&1  &\delta>\delta_{c}.\\
\end{aligned}
\right.
\end{equation}
There is theoretically a small remaining discontinuity at $\delta_i=\delta_c $ and $\delta_i=0 $, denoted as $\beta$, which is\footnote{In practice, the sky being pixelised, any mask has unavoidable discontinuities between adjacent pixels.}:
\begin{equation}
\beta=\frac{1}{2}-\frac{1}{2}{\rm erf}\bigg(\frac{\frac{\delta_c}{2}}{\sqrt{2}\sigma }\bigg).
\end{equation}
Here we set $ \delta_c=1^\circ$ and $\beta=10^{-4}$, the latter defining the value of $\sigma$.  The resulting mask is displayed in Figure~\ref{mask_planck}.

We then use with that particular mask the method proposed by 
Smith (2006) and Smith \& Zaldarriaga (2007), to compute pure-pseudo multipoles coefficients for E and B, avoiding
the ``$E$-to-$B$'' leakage. 

\subsection{Analysis pipelines}

We apply the ABS method to two different cases. In the first case, we analyze full-sky maps with no masking. This avoids the $E$-$B$ confusion problem, at the price of strong foreground contamination in regions close to the Galactic plane. The second case considers partial-sky observations, after applying the mask displayed in Figure~\ref{mask_planck}, with reduced foreground contamination but unavoidable $E$ to $B$ leakage.

In the first case, ABS is applied in the spectral domain after a full-sky analysis of the input maps into harmonic coefficients. The covariance matrix of the observations, $\mathcal{D}^{\rm obs}$, is computed $\ell$ by $\ell$, and the power spectrum is computed following the ABS method as described in Section \ref{sec:abs}.

In the second case, we must modify the procedure to take into account the impact of the mask. We follow the steps below:

\begin{enumerate}[(A)]
\item We smooth all maps to a common resolution of $28.3'$;
\item We apply the apodized Planck 2015 component separation common polarization mask (UPB77 with $f_{sky}=74.68\%$) to the final $Q$ and $U$ simulations: CMB $+$ foregrounds $+$ noise for each frequency band;
\item We reconstruct the pure $B$-mode pseudo-multipoles for every frequency map, following the method of Smith and Zaldarriaga;
\item We then extract the final CMB $B$-mode spectrum using the ABS method;
\item We repeat the procedure for 50 independent realizations of the instrumental noise, keeping the CMB signal and the foregrounds fixed; 
\item We calculate the mean and the standard deviation of the estimated  $B$-mode power spectrum based on the results from steps (C) and (D).
\end{enumerate}

\section{Results}
\label{results}

We now present and discuss the results obtained with the ABS method in the recovery of the CMB $E$- and $B$-mode power spectra, based on the simulated polarization maps made by the fiducial survey specification given int Table~\ref{tab:exp}. Means and  associated statistical errors are computed from the average of 50 simulations with independent noise realizations.

\subsection{The full-sky case with two foreground components}\label{sect:full_sky}

\begin{figure}[!htpb]
\centering
\includegraphics[width=3.2in] {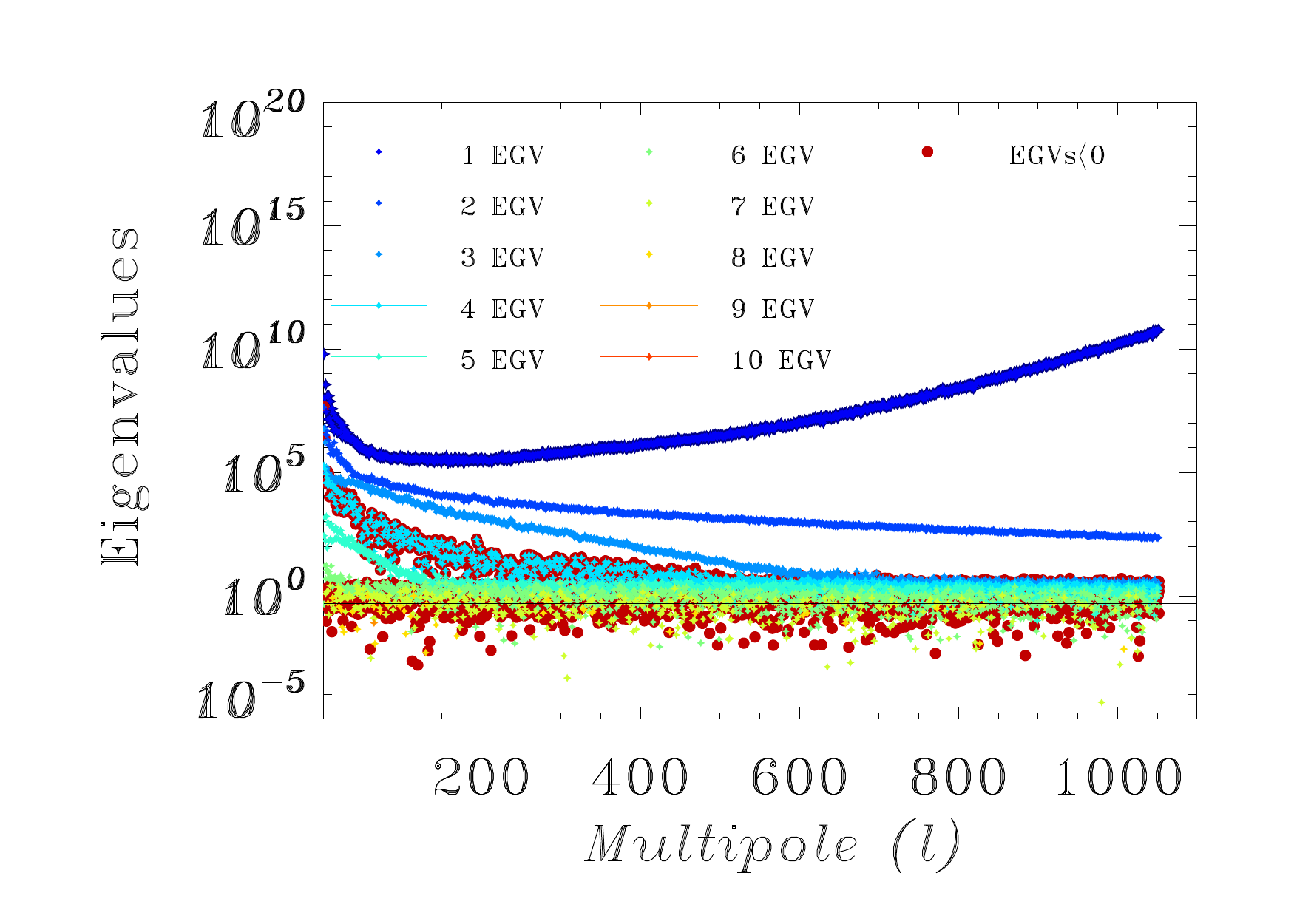}
\includegraphics[width=3.2in] {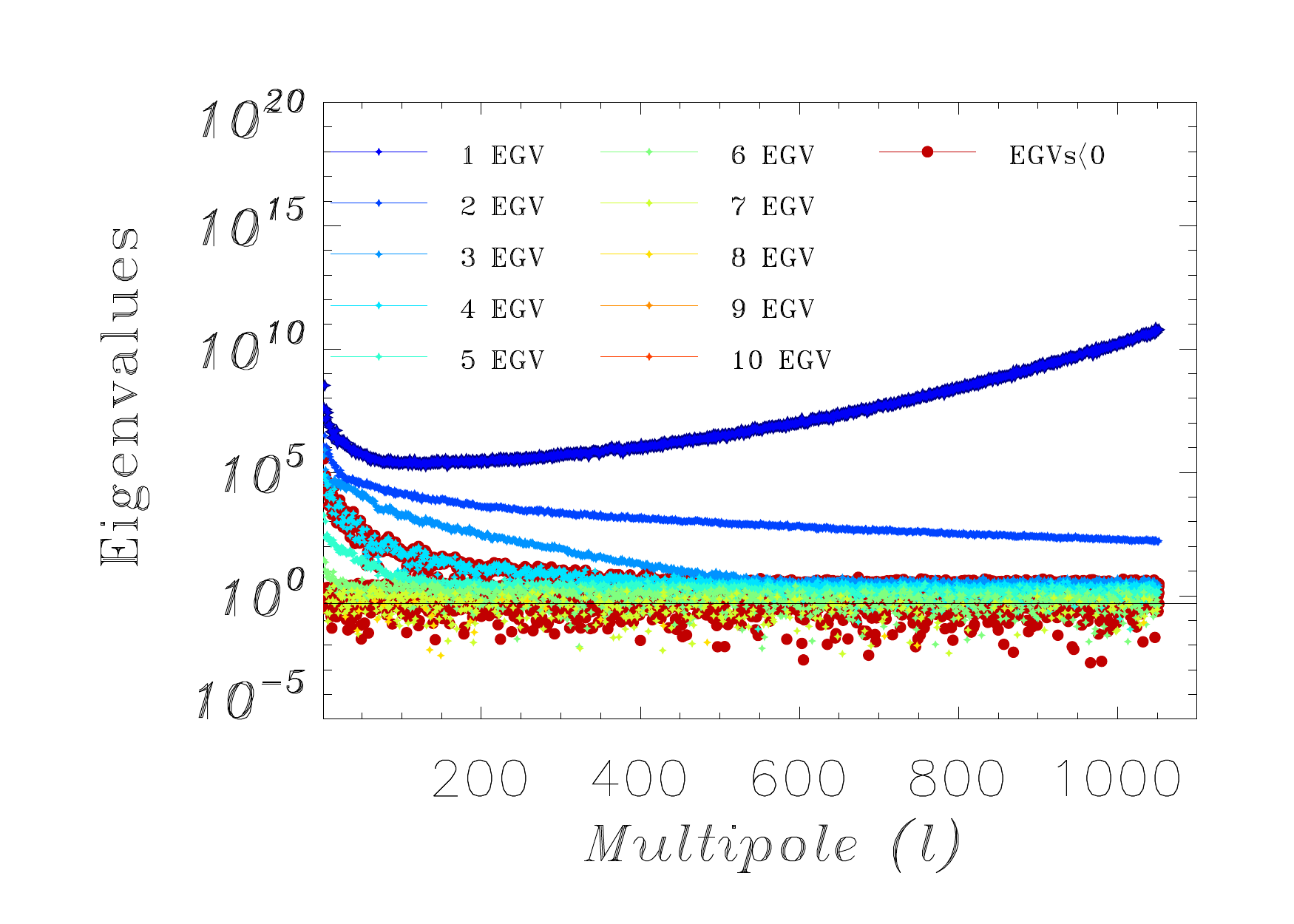}
\caption{Eigenvalues (labelled as EGV above) of $\mathcal{D}^{\rm obs}_{ij}(\ell)$ for each of the 10 frequency bands (represented in different symbol colors), considering CMB, foregrounds (synchrotron and thermal dust) and one noise realization. Eigenvalues from the $E$-mode (upper panel) and $B$-mode (lower panel) are shown. The threshold $\tilde{\lambda}_{\rm cut}=1/2$ is displayed as an horizontal solid black line. Note that $\tilde{\mathcal{D}}^{\rm obs}_{ij}(\ell)$ is not strictly positive: the subspace associated to the lowest eigenvalues is dominated by noise, and hence corresponding eigenvalues are centered around zero after the subtraction of the noise covariance from $\tilde{\mathcal{D}}^{\rm obs}_{ij}(\ell)$. The eigenvalues, $\tilde{\lambda}_\mu$, are shown in absolute value,  negative eigenvalues being displayed as red dots.} 
\label{egv_full} 
\end{figure}

The distribution of eigenvalues in the full-sky case, as a function of multipole $\ell$, is displayed for both $E$ and $B$ modes in Figure~\ref{egv_full}. There are $10$ eigenvalues for each multipole, as $\mathcal{D}^{\rm obs}_{ij}(\ell)$ is the covariance of observations in 10 frequency channels. In spite of the fact that there are only two foreground components and one CMB, we see that at low multipoles there are more than three significant eigenvalues above threshold. For all $\ell$, however, the lowest eigenvalues are scattered around or below the threshold, showing that the simulated microwave sky is completely decomposed into those eigenmodes.

\begin{figure}[!htpb]
\centering
\includegraphics[width=3.5in] {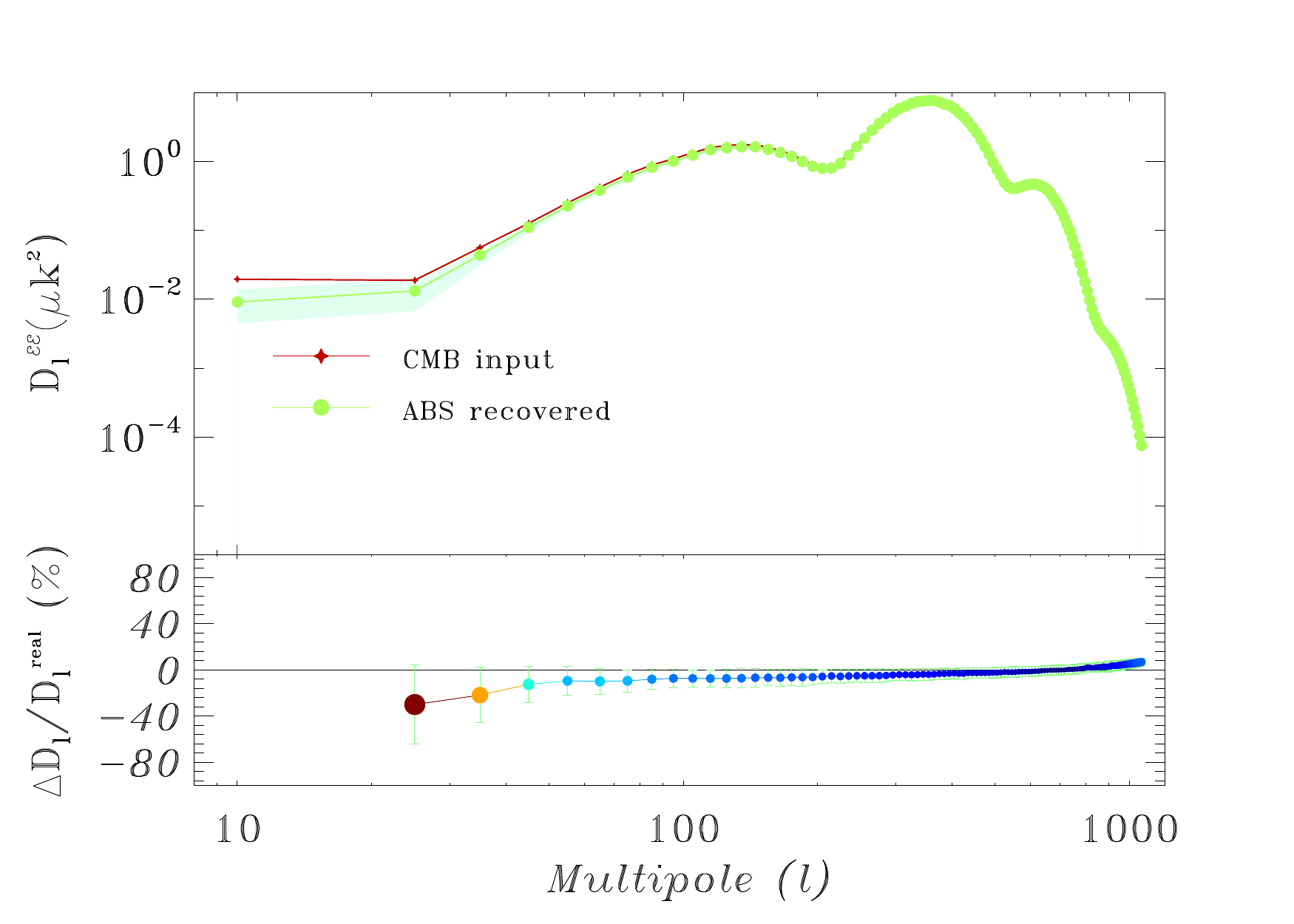}
\caption{Full-sky binned CMB $E$-mode power spectrum estimated with ABS, for $r=0$, and a Gaussian beam of FWHM=$28.3'$. The red curve corresponds to the simulated CMB $E$-mode spectrum, while the green curve is the ABS estimate. The associated 1-$\sigma$ statistical errors are also shown as a light green shadow region, computed from 50 simulations with independent realizations of the instrumental noise. The lower part of the figure shows the relative error in the estimated power spectrum, ${\mathcal{D}}^{\rm rec}_\ell / \mathcal{D}^{\rm true}_\ell -1$, in per cents. The color and size of the symbols emphasize deviations from 0\% error with respect to the input spectrum.} 
\label{full_r0E} 
\end{figure}

The result for the full-sky analysis of $E$ modes is shown in Figure \ref{full_r0E}, which compares the ABS result with the power spectrum directly obtained from the pure CMB $Q$ and $U$ maps without foregrounds and noise, {\it but smoothed to $28.3'$}  (dubbed as the ``true'' power spectrum) for $\ell_{ max}=1050$. We also show the relative error, in percentage, calculated as: $\widehat{\mathcal{D}}^{\rm cmb}_\ell / \mathcal{D}^{\rm real}_\ell -1$ where $\widehat{\mathcal{D}}^{\rm cmb}_\ell$ is the estimator described in Equation~\ref{eq:abs}. 

Although the analysis is performed $\ell$ by $\ell$, we show power spectra after binning in $\ell$, with bin width 
$\Delta \ell = 20$ for the first bin, and $\Delta \ell = 10$ for the other bins. 
Overall, all the estimated band-powers are within error. However, we see a tendency for low-$\ell$ estimates to be lower than the input spectrum, and high-$ell$ estimates to be somewhat above the spectrum. The former effect is likely to be due to empirical correlations between CMB and contaminants (foregrounds and noise). The slight excess power at high ell might be due to the  thresholding bias discussed in section \ref{sec:biases}, and illustrated on a simple case in Appendix \ref{appendix:special-case}.

\begin{figure}[!htpb]
\centering
\includegraphics[width=3.5in] {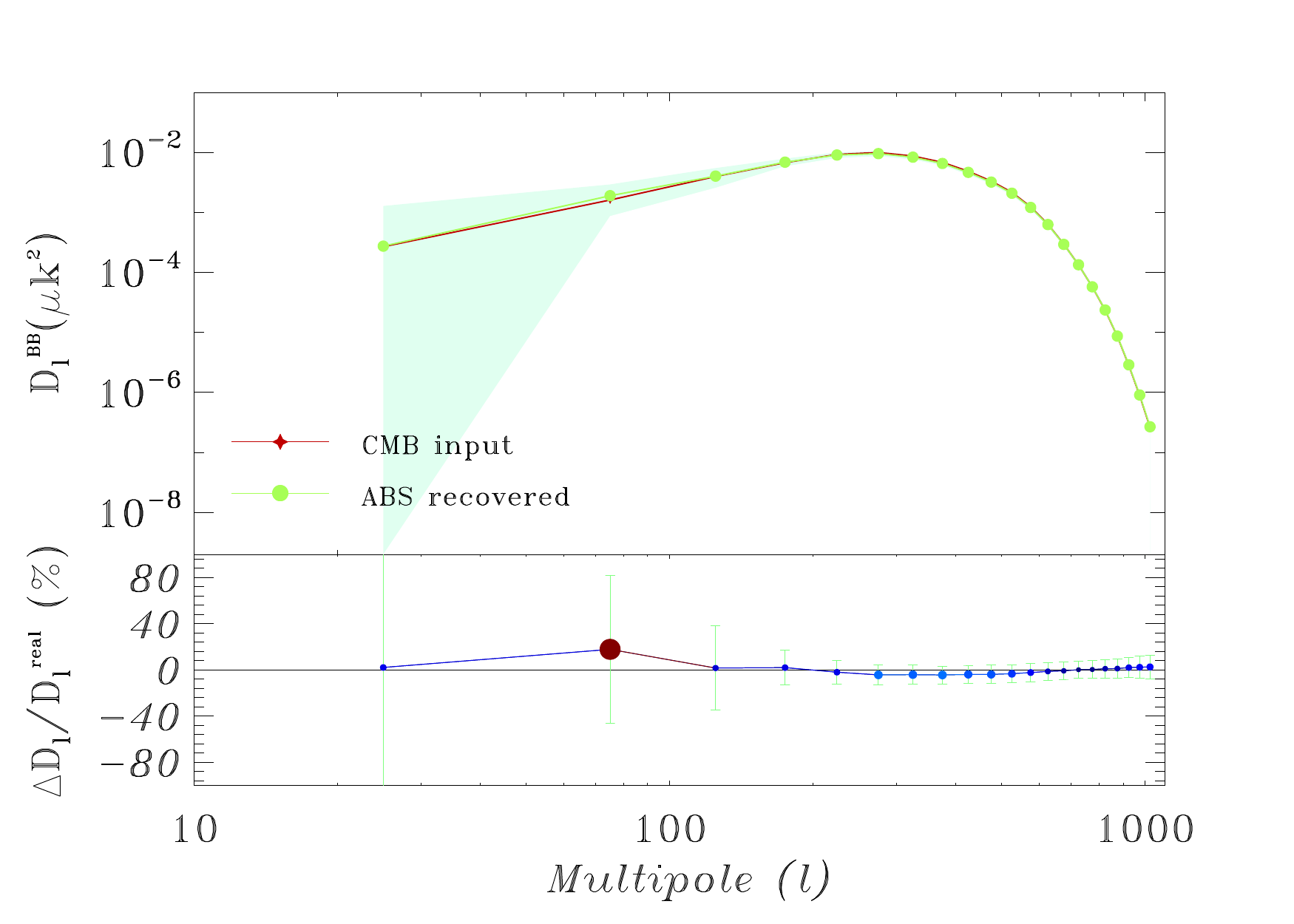}
\includegraphics[width=3.5in] {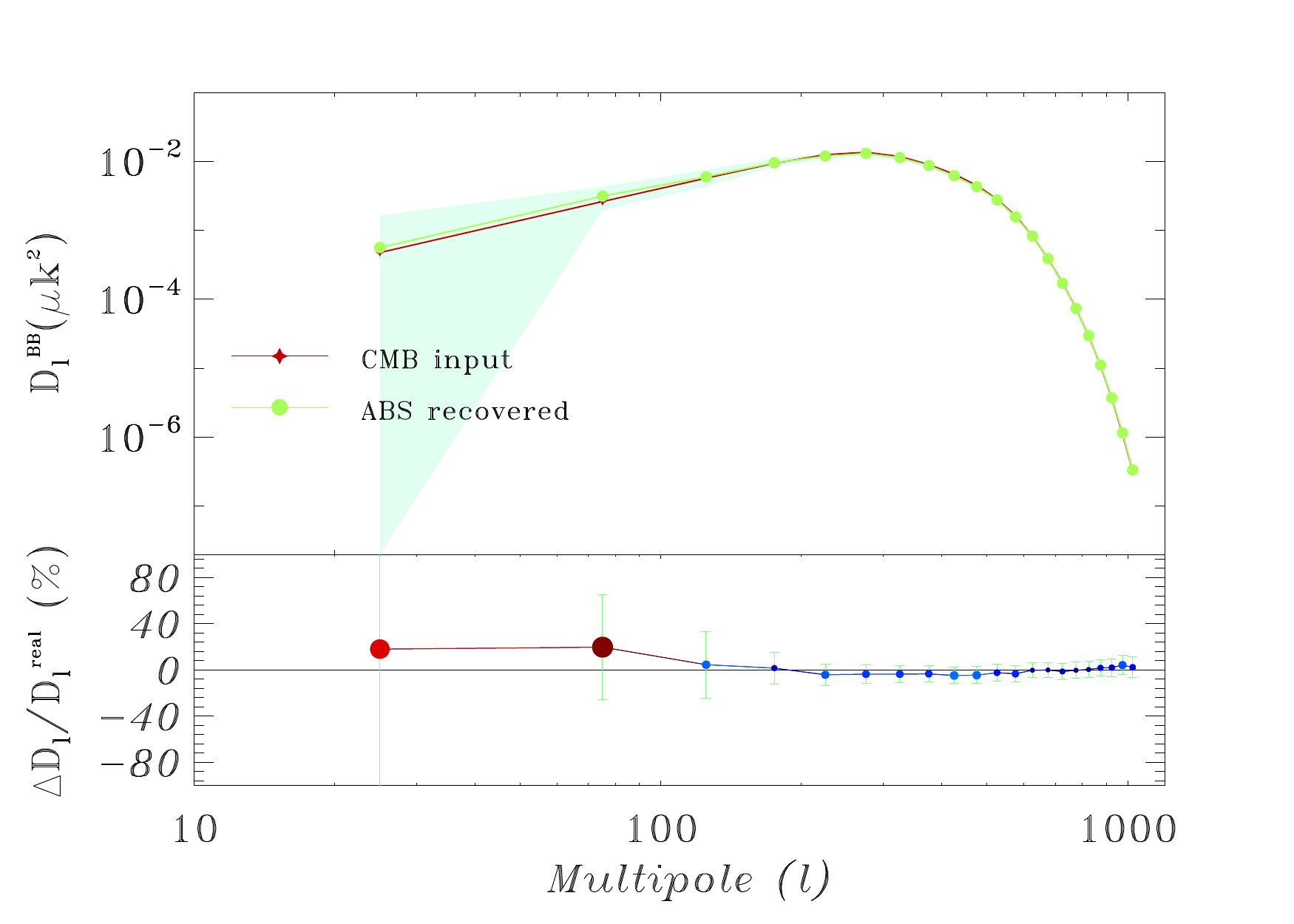}
\includegraphics[width=3.5in] {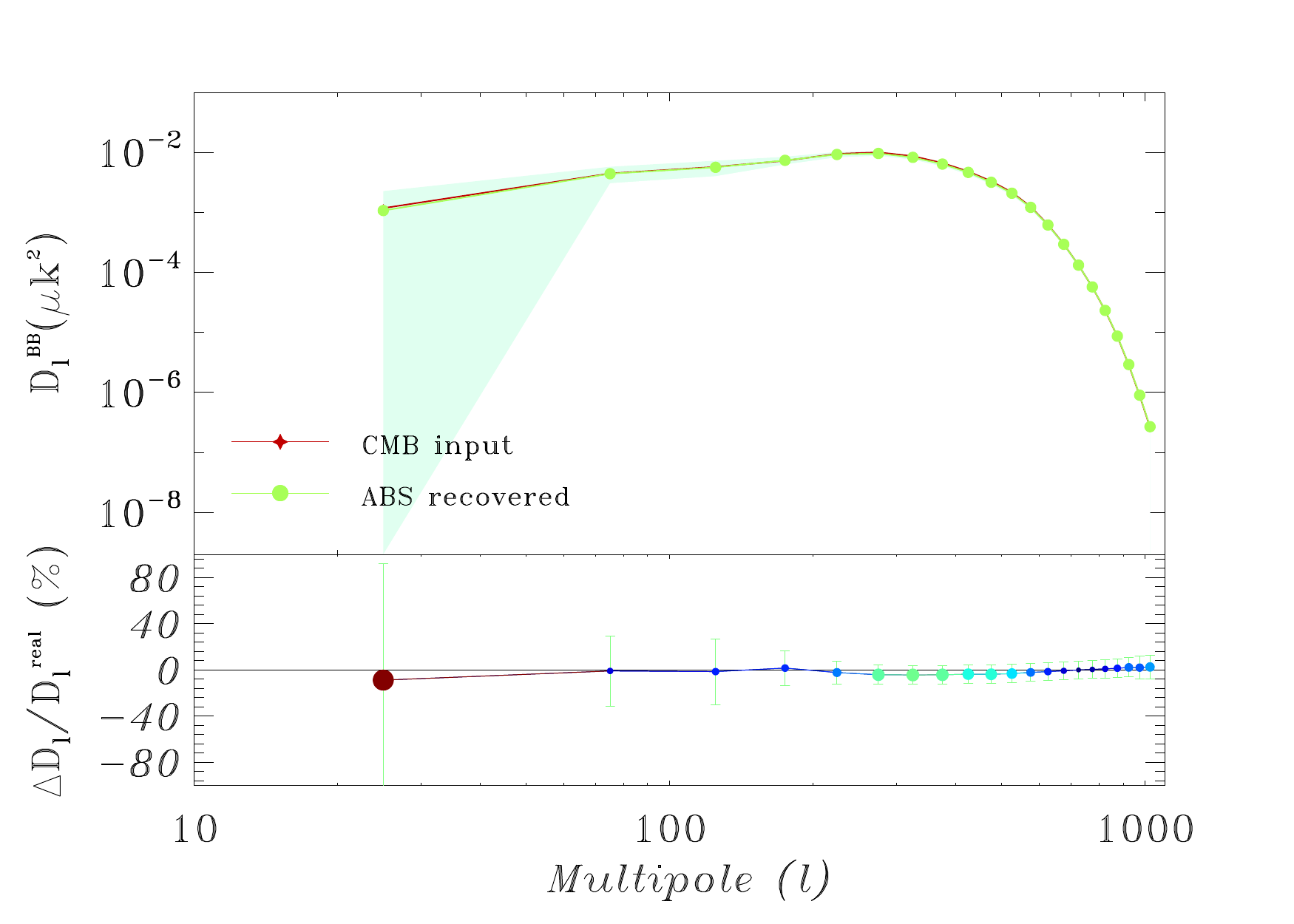}
\caption{Full-sky $B$-mode spectrum for $r$=0 (top), $r$=0.01 (middle), and $r$=0.05 (bottom). The red curves correspond to the simulated CMB $B$-mode spectrum (including lensing $B$-modes, while the green curves are the ABS estimate. The lower parts of the figures show the relative error in the estimated power spectra. The color and size of the symbols emphasize deviations from 0\% error with respect to the input spectrum.
} 
\label{fig:full-sky-B} 
\end{figure}

Results for $B$ modes are displayed in Figure \ref{fig:full-sky-B}. For the fainter $B$-modes, we use a bin width of $\Delta \ell = 50$ for displaying the results. For $r=0$ and $r=0.01$, we find that the difference between the recovered spectrum and the input spectrum is below $20\%$ for the full multipole range, even though the 1-$\sigma$ confidence level for the reconstructed $B$-mode power spectrum is quite large at the low-$\ell$ region. In the case of $r=0.05$, the relative performance is better than 10\% for all multipoles. In all cases, for $\ell>150$, the recovered power spectrum differs from the input spectrum by less than $5\%$. At the very lowest $\ell$s, however, the errors and biases become too large for a successful estimate of the $B$-mode power spectrum. Considering the relative level of foreground contamination as compared to the CMB $B$ modes. This is, however, not a big surprise.

\begin{figure}[!htpb]
\centering
\includegraphics[width=3.5in] {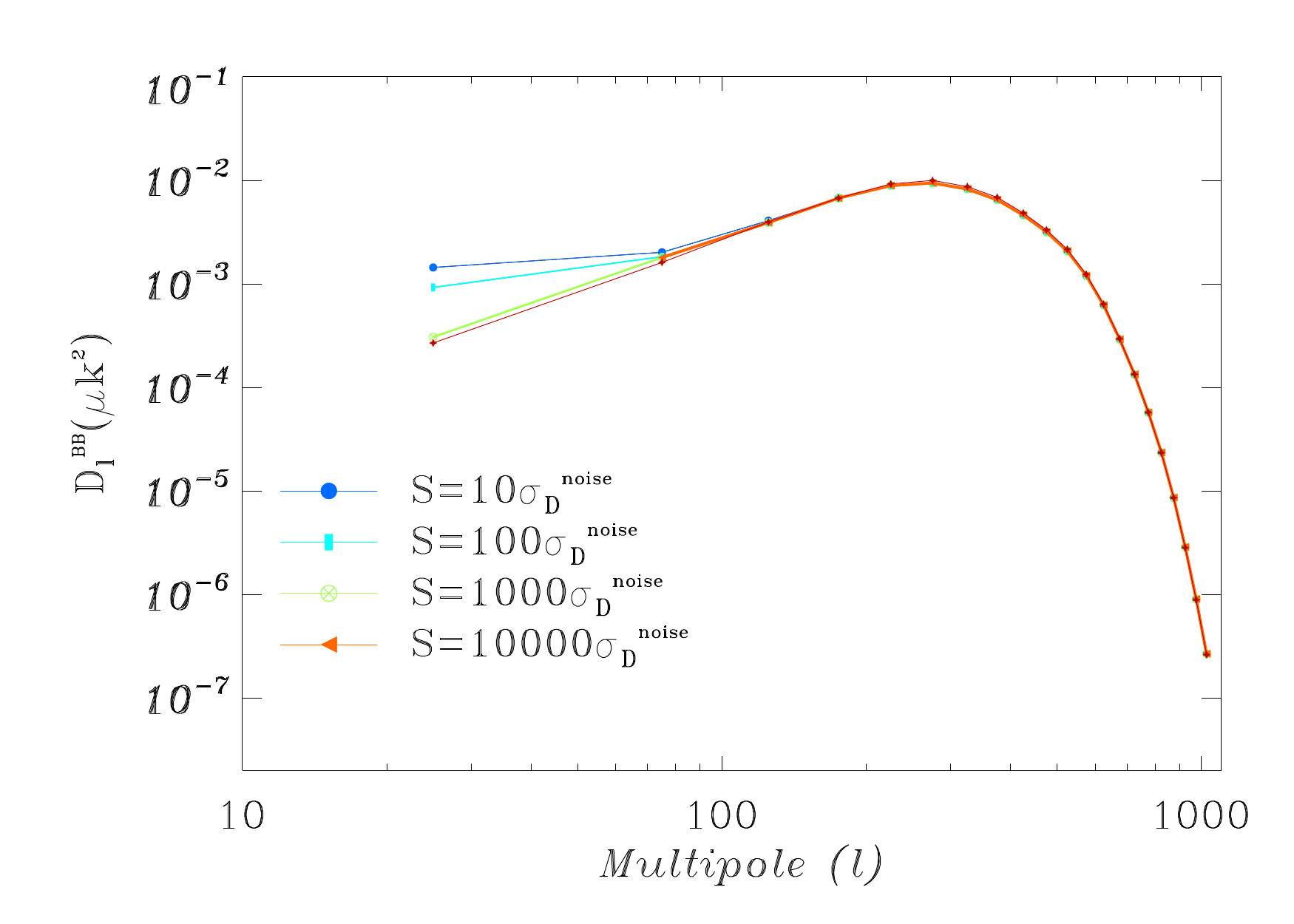}
\caption{Full-sky reconstructed $B$-mode power spectrum for $r=0$ with different values for the shift parameter $\mathcal{S}$.} 
\label{fig:full_r0-different-S} 
\end{figure}

In Figure \ref{fig:full_r0-different-S}, we show the change in the $B$-mode spectrum for different values of the shift parameter $\mathcal{S}$ introduced in Equation \ref{eq:absshift}. We show the recovered spectrum for $r = 0$ using four different values of $10\sigma_{\mathcal{D}}^{\rm noise},100\sigma_{\mathcal{D}}^{\rm noise},1000\sigma_{\mathcal{D}}^{\rm noise},10000\sigma_{\mathcal{D}}^{\rm noise}$. 
This shows the impact of the choice of $\mathcal{S}$ on the power spectrum estimation with ABS. An illustration of how this choice may impact the estimator in a simple case can be found in Appendix \ref{appendix:special-case}.
As mentioned in Section \ref{sec:biases}, on the basis of this investigation, we set the nominal value of the shift parameter to $\mathcal{S} = 1000\sigma_{\mathcal{D}}^{\rm noise}$.

\begin{figure}[!htpb]
\centering
\includegraphics[width=3.5in] {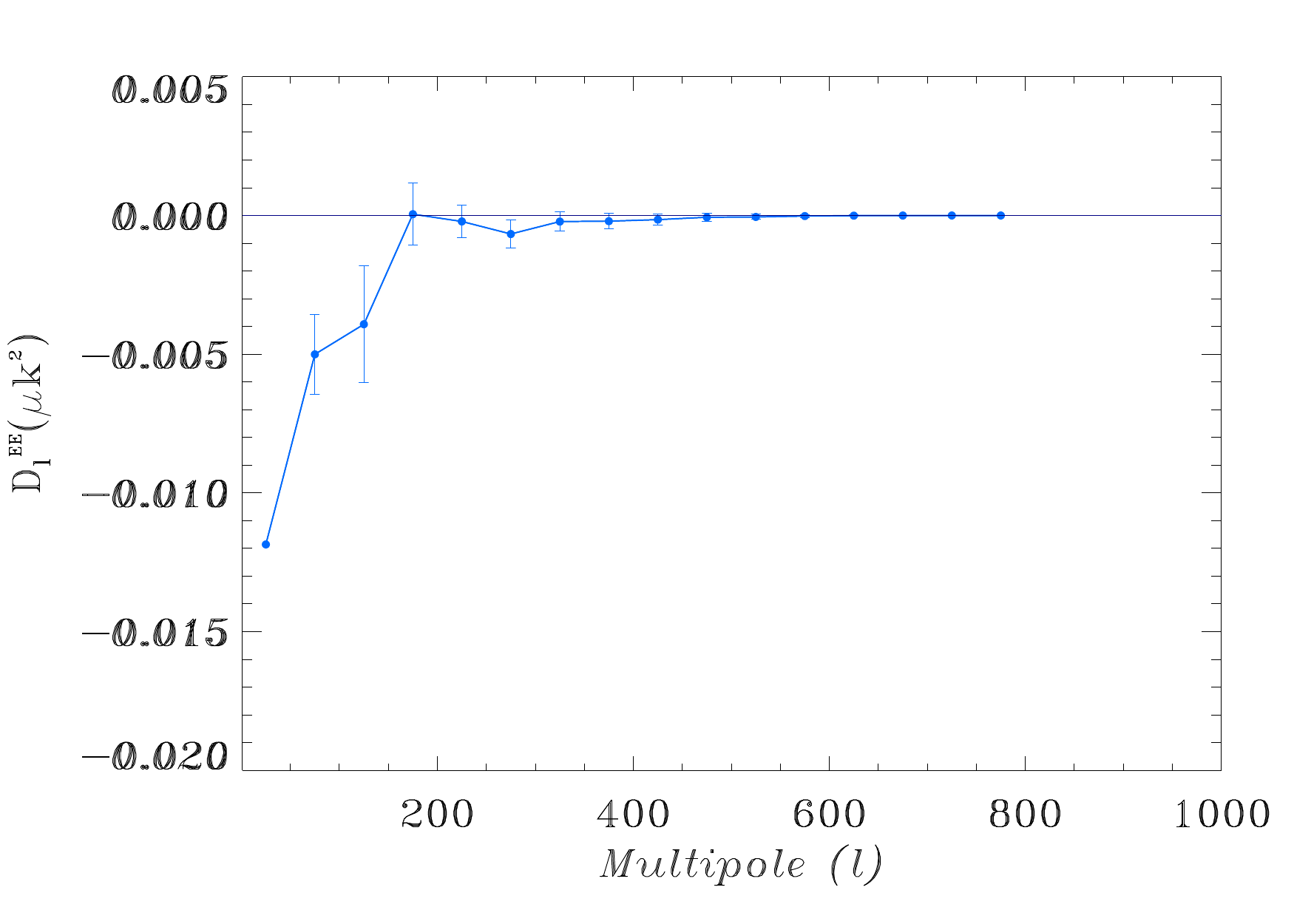}
\includegraphics[width=3.5in] {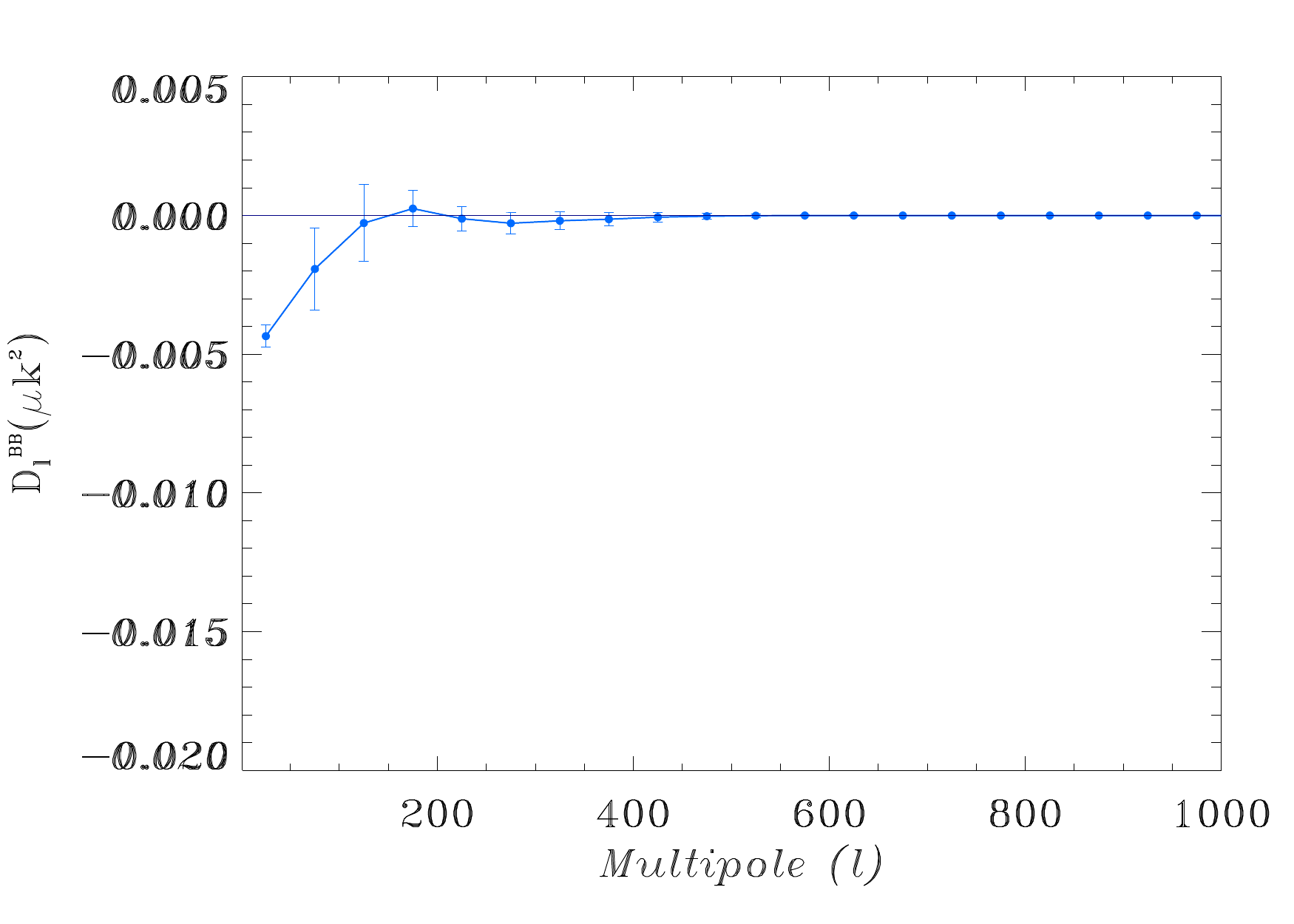}
\caption{The null result (foreground + noise only) for the $E$-mode (upper panel) and $B$-mode (lower panel).} 
\label{full_S_null} 
\end{figure}

Finally, we show in Figure \ref{full_S_null} the result of a null test, in which we analyse with the exact same pipeline (and same shift parameter and threshold $\lambda_{\rm cut}$) a simulation with no CMB signal in the $Q$ and $U$ maps, but only foregrounds and Gaussian noise. The null test is important to assess the limitations and possible biases of our estimator, especially for detecting the extremely faint primordial $B$-mode polarization signal. We find a small, but significant underestimation in the null spectra below $\ell=150$, in agreement with the other results shown in this section.

Overall, the performance of ABS for power spectrum estimation in those tests is quite satisfactory, taking into account the fact that those estimations have been done \emph{on full-sky simulations}, with foreground contamination power exceeding the CMB power by orders of magnitude, as shown in Figure \ref{fig_PS}. However, although mostly compatible with error bars, $E$-mode power spectrum estimates in low-$\ell$ bins are systematically lower than the input spectrum. This is probably due either to the finite sample bias or the thresholding bias discussed in section~\ref{sec:biases}, or a combination of both, and is confirmed by the null-case analysis, in which the recovered spectrum is systematically negative for low-$\ell$ bins. Reducing these biases would be possible with a careful choice of $\ell$-bins to compute the original covariance of the observations (i.e. binning \emph{before} ABS estimation, to increase the statistics) and of $\lambda_{\rm cut}$. We postpone this optimization to future work. For $B$ modes, even though the ABS result seems promising, improvement of the power spectrum reconstruction at this low multipole range is necessary for a detection of the primordial gravitational wave imprint in the CMB polarization signal. Resolving the spectrum for low $\ell's$, especially for $\ell<40$, is challenging if we consider the ABS foreground separation method in a full-sky approach. We must then consider an analysis in which we mask the most foreground-contaminated sky regions.

Similar results in the case of a foreground model that includes polarized anomalous microwave emission (AME) are shown in Appendix \ref{app:AME}.

\subsection{The partial-sky case}\label{sect:partial_sky}

Using the same CMB, synchrotron and dust simulations, and noise realizations as in the previous section, we now consider the $E/B$ decomposition in partial sky for each analyzed frequency band, using both the ABS foreground cleaning approach and the pseudo-power spectrum reconstruction through the Smith and Zaldarriaga method. 



\begin{figure}[!htpb]
\centering
\includegraphics[width=3.5in] {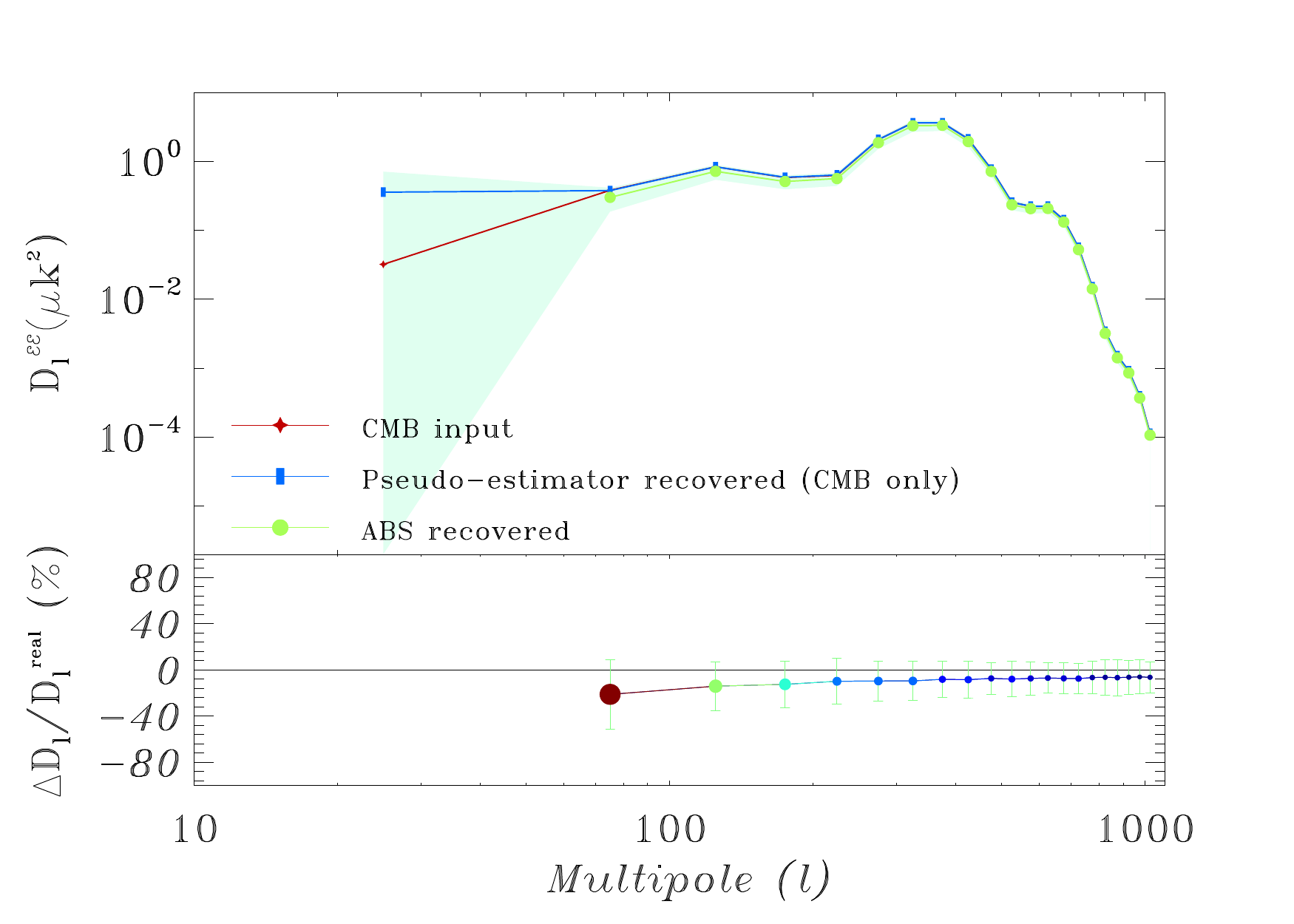}
\caption{Part-sky CMB $E$-mode power spectrum (binned with $\Delta l=50$ bands) estimated with  ABS (green curve). The red curve is the CMB $E$-mode input power spectrum. The blue curve, which shows the pseudo-power spectrum computed directly from the CMB sky realization without noise/foregrounds, using the Smith and Zaldarriaga $E/B$ separation method, illustrates the uncertainties inherent to the correction of the $E$-$B$ mixing. The 1-$\sigma$ statistical errors, computed from 50 simulations with independent realizations of the instrumental noise, are shown as the light green shadow region. The relative error, $\mathcal{D}^{\rm rec}_\ell / \mathcal{D}^{\rm real}_\ell -1$, is shown in percentage level (the error in the first bin, which is much larger than the y-axis scale, is not displayed). The symbol colors and sizes illustrate deviations from 0\% in respect to the ``true''  spectrum.}
\label{partial_r0E} 
\end{figure}

\begin{figure}[!htpb]
\centering
\includegraphics[width=3.5in] {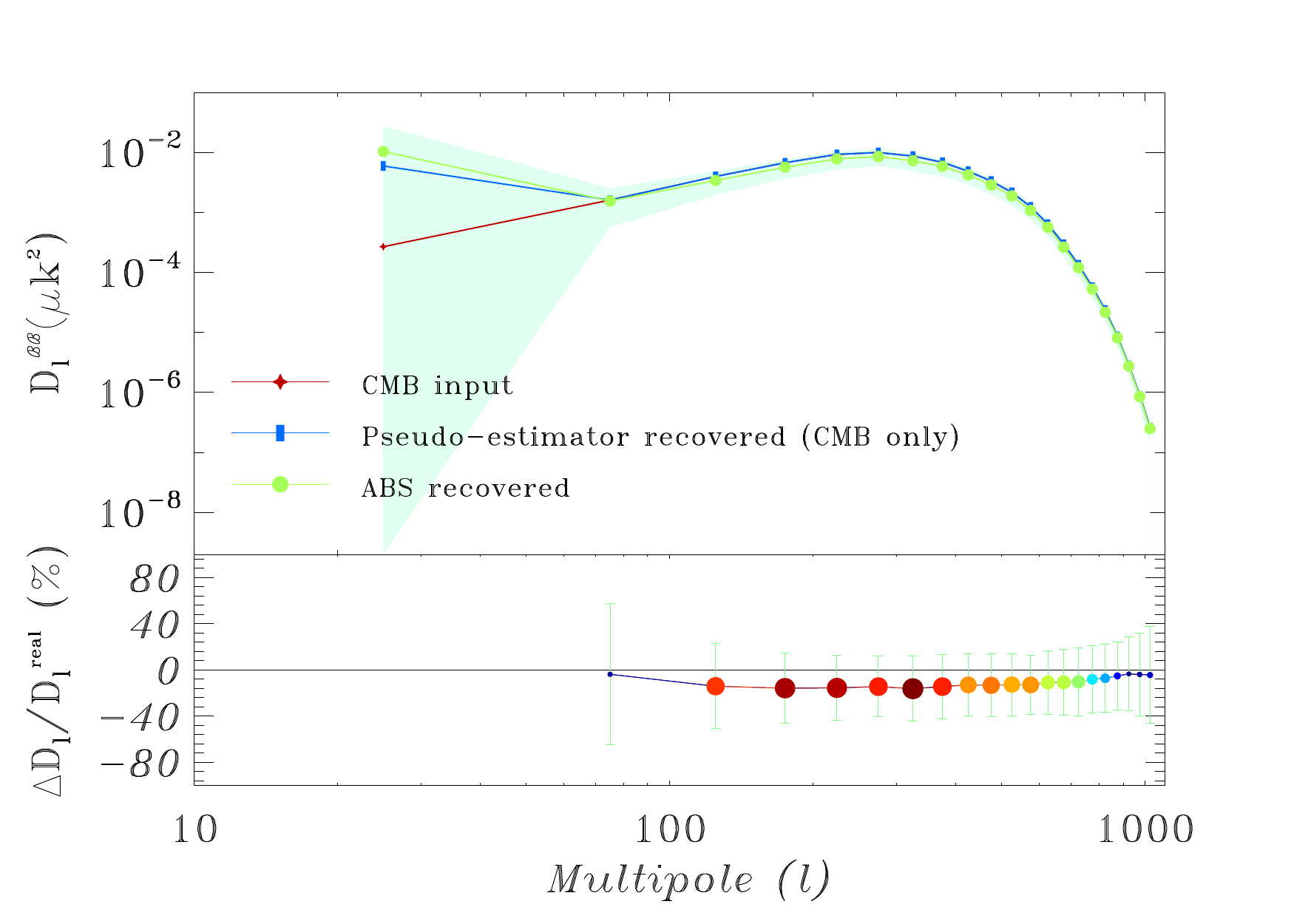}
\includegraphics[width=3.5in] {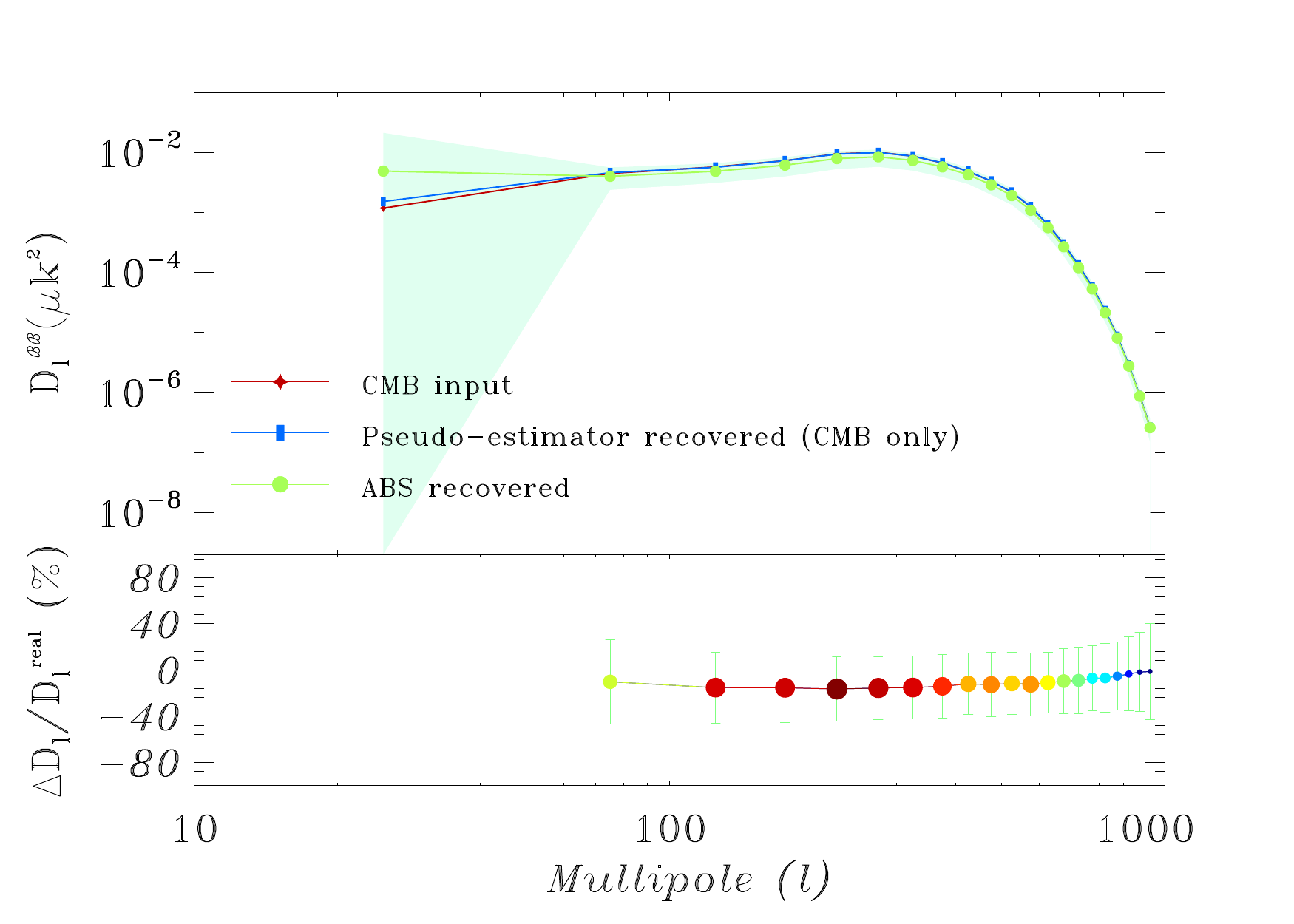}
\caption{Same as Fig.~\ref{partial_r0E} but for the $B$-mode power spectrum, for  $r=0$ (upper panel) and $r=0.05$ (lower panel).}
\label{fig:part-sky-B} 
\end{figure}

The result for $E$-modes (computed for the $r = 0$  case) is shown in Figure \ref{partial_r0E}. The performance of the ABS method in recovering the $E$-mode and $B$-mode power spectra for $30 \leq \ell \leq 1050$ is comparable with the full-sky case. The difference between the recovered power spectrum and the ``true'' one is below $17\%$ for the mentioned multipole range, again with a slight underestimation at low multipoles. For the $B$-modes, shown in Figure \ref{fig:part-sky-B}, we find the relative error to be less than $21\%$ for $30 \leq \ell \leq 1050$, and less than $15\%$ for $100 \leq \ell \leq 1050$. 


For smaller multipoles (and especially in the first bin), the determination is not very accurate. The  uncertainties here originate from \emph{both} the ABS method and the $E$-$B$ mixing correction, the performance of which deteriorates rapidly below $\ell=50$ in our case. In the same way, the results of the $E$-$B$ mixing correction deteriorate at high $\ell$. A more detailed investigation of the effect, with no foreground contamination, can be found in Appendix~\ref{app:E-B}. 

We do not try improving on the $E$-$B$ mixing correction with more sophisticated approaches for polarization reconstruction on partial sky here. Options are discussed, for instance, by ~\citet{ferte2013,ebmixture9,ebmixture10,ghosh2020}. We leave such extensions of our investigations to future work. We simply note that for partial-sky surveys it is important, especially in the low-$\ell$ regime, to carefully perform an analysis with optimized power spectrum estimation approaches to extract the primordial $B$-modes. This is particularly critical here, as the ABS method post-processes the covariance of the observations, and small errors in some of the entries of the covariance matrix (i.e., auto and cross spectra between the different channels) can propagate in unpredictable ways into the ABS eigendecomposition.

\begin{figure}[!htpb]
\centering
\includegraphics[width=3.5in] {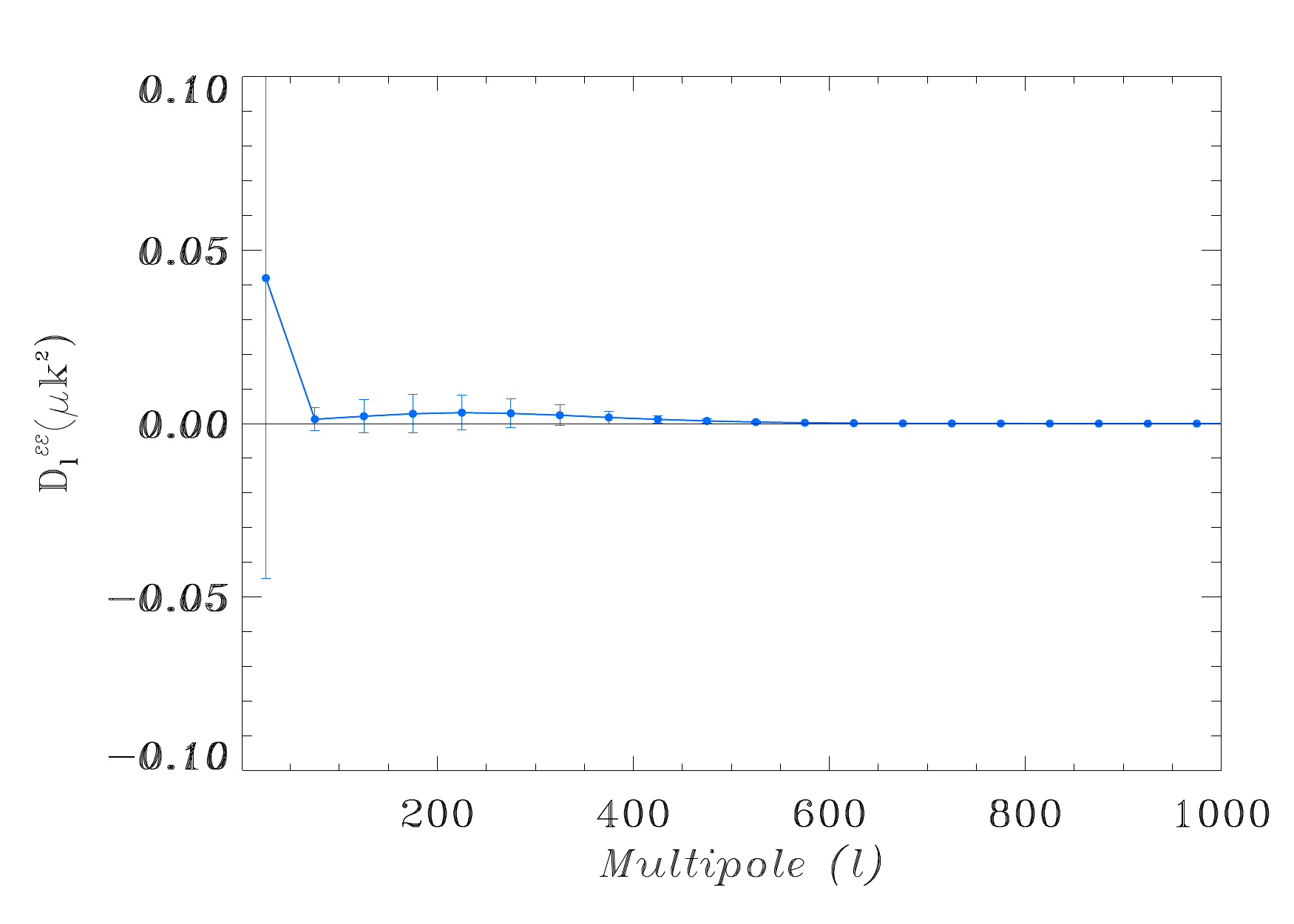}
\includegraphics[width=3.5in] {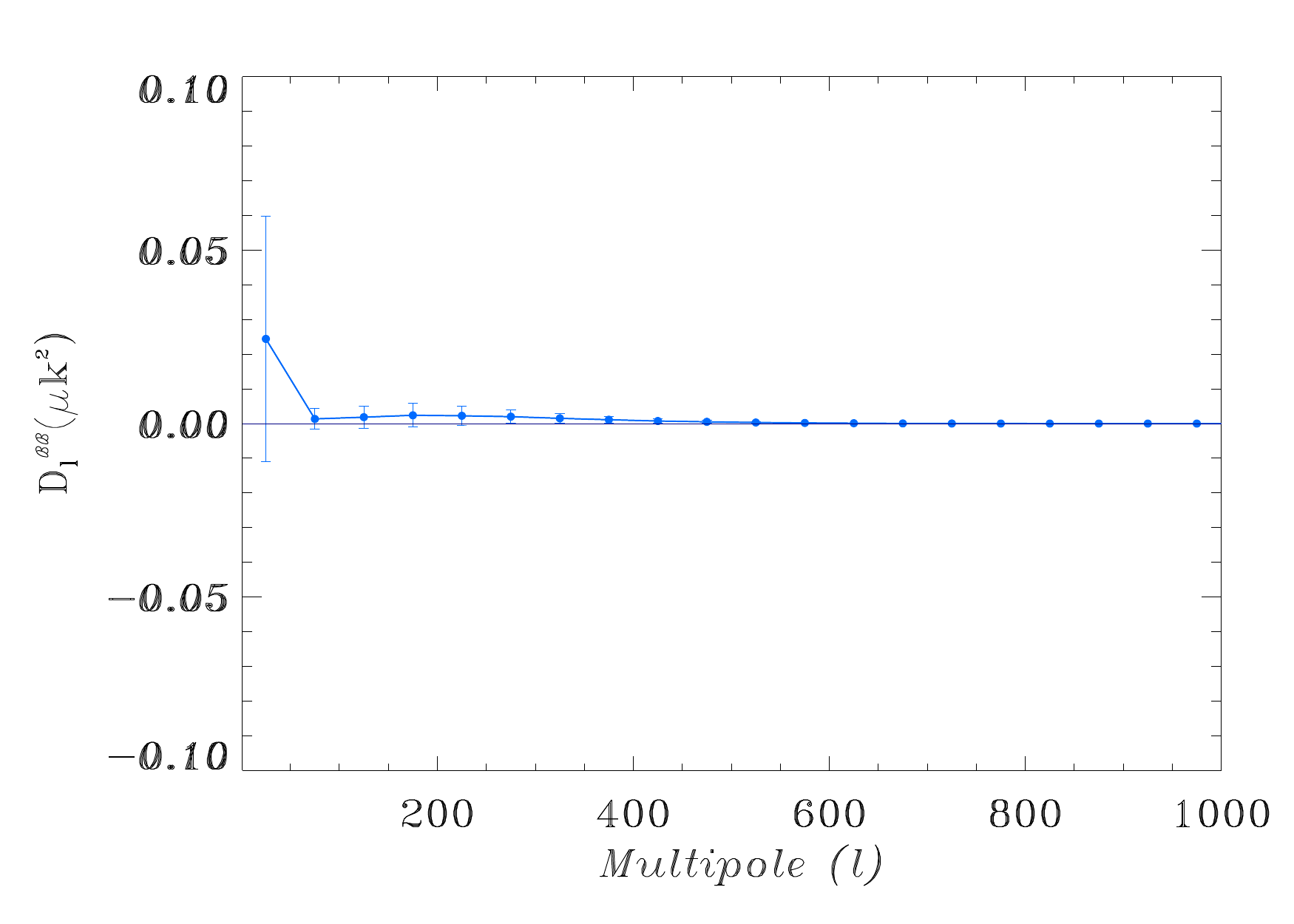}
\caption{The null result (foreground + noise only) for the $E$-mode (upper panel) and $B$-mode (lower panel).} 
\label{partial_S_null} 
\end{figure}

Finally, in Figure \ref{partial_S_null}, we see the results for the null test in the partial-sky analysis. The biases at low $\ell$ that were seen on the full-sky case are not visible, but this is probably mostly by reason of the large error bar induced by the $E$-$B$ mixing correction.


\section{Conclusions}
\label{sect:con}


In this study, we have tested the performance of the ABS method to determine CMB $E$-mode and $B$-mode power spectra from multi-frequency observations contaminated by both foreground emission and instrumental noise. The ABS estimator has been applied to simulated maps as could be observed by a hypothetical future experiment with 10-frequency bands in the range of 30 to 321 GHz. Taking into account 50 independent noise realizations, with one simulated CMB signal and two foreground components, we have analyzed a set of 50 $Q$ and $U$ maps to evaluate the performance of the recovery of the $E$- and $B$-mode power spectra.  We find that the ABS method is able to estimate both CMB $E$- and $B$-mode power spectra within 1-$\sigma$ error bars at most scales ($50 \leq \ell \leq 1050$). In the case of full sky, we calculate the ABS estimator for tensor-to-scalar ratios $r=0$, $r=0.01$, and $r=0.05$. The polarized $E$-mode power spectrum can be recovered within $20\%$ for $\ell>30$, reaching an agreement with the input spectrum of  $5\%$ compared with the input spectra for $\ell>150$. For the $B-$mode spectrum, considering $\Delta \ell=50$, we found that the difference between the recovered and the input spectrum is also below $20\%$ for the full multipole range in every tested case. However, it is important to point out that in all the recovered $B-$mode spectra, the 1$\sigma$ confidence interval is large for low $\ell$'s, and the method therefore provides poor information with respect to the detectability of low values of~$r$. 

Even though our results show that a disentanglement of the primordial $B$-mode from noise and foregrounds is challenging with ABS on the full sky in the observational configuration that was tested here, the lensing signal can be recovered with excellent accuracy, in spite of the strong contamination that comes from foreground emission in the Galactic plane.

In the case of a partial-sky analysis, combining the ABS together with the Smith and Zaldarriaga method to reduce the $E$-to-$B$ leakage, we find that the recovery of the $E$ and $B$ power spectra, for most of the scales ($30 \leq \ell \leq 1050$), has a relative error below $21\%$, for both $r=0$ and $r=0.05$. However, the results are not very satisfactory in the low-$\ell$ region (for both $E$ and $B$). Contrarily to methods that first construct a foreground-cleaned CMB map and then estimate the power spectrum of that map, the ABS method does not allow for masking contaminated regions after foreground cleaning and before power spectrum estimation. It is plausible that an interaction between errors in the calculation of pseudo spectra and the ABS eigenvalue decomposition is limiting the perfomance of the combination of these two tools. We leave further investigations of those effects to future work.



%

\section*{Acknowledgments}
This work was supported by the National Science Foundation of China (11621303, 11653003, 11773021, 11433001), the National Basic Research Program of China (2015CB85701, 2013CB834900), and the National Key R\&D Program of China (2018YFA0404601). Larissa Santos, Wen Zhao, Shamik Ghosh and Jiming Chen are sup\-ported by NSFC Grants 11773028, 11603020, 11633001, 11173021, 11322324, 11653002,  11421303 and 11903030, by the project of Knowledge Innovation Program of Chinese Academy of Science, the Fundamental Research Funds for the Central Universities, and the Strategic Priority Research Program of the Chinese Academy of Sciences Grant no. XDB23010200. T.V. acknowledges  CNPq  Grant  308876/2014-8.

\appendix 

\section{Including AME polarized emission}
\label{app:AME}

We repeat our full-sky analysis for a foreground model that includes polarized anomalous microwave emission (AME). Although it is not expected that AME be strongly polarized, \cite{Remazeilles2016} showed it can bias the derived value of the tensor-to-scalar ratio by non-negligible amounts for satellite missions if an 1\% level of polarized AME is neglected. We use the AME polarized model (model 2) from the ~\texttt{PySM} model~\citep{2017MNRAS.469.2821T}. 

\begin{figure}[!htpb]
\centering
\includegraphics[width=3.5in] {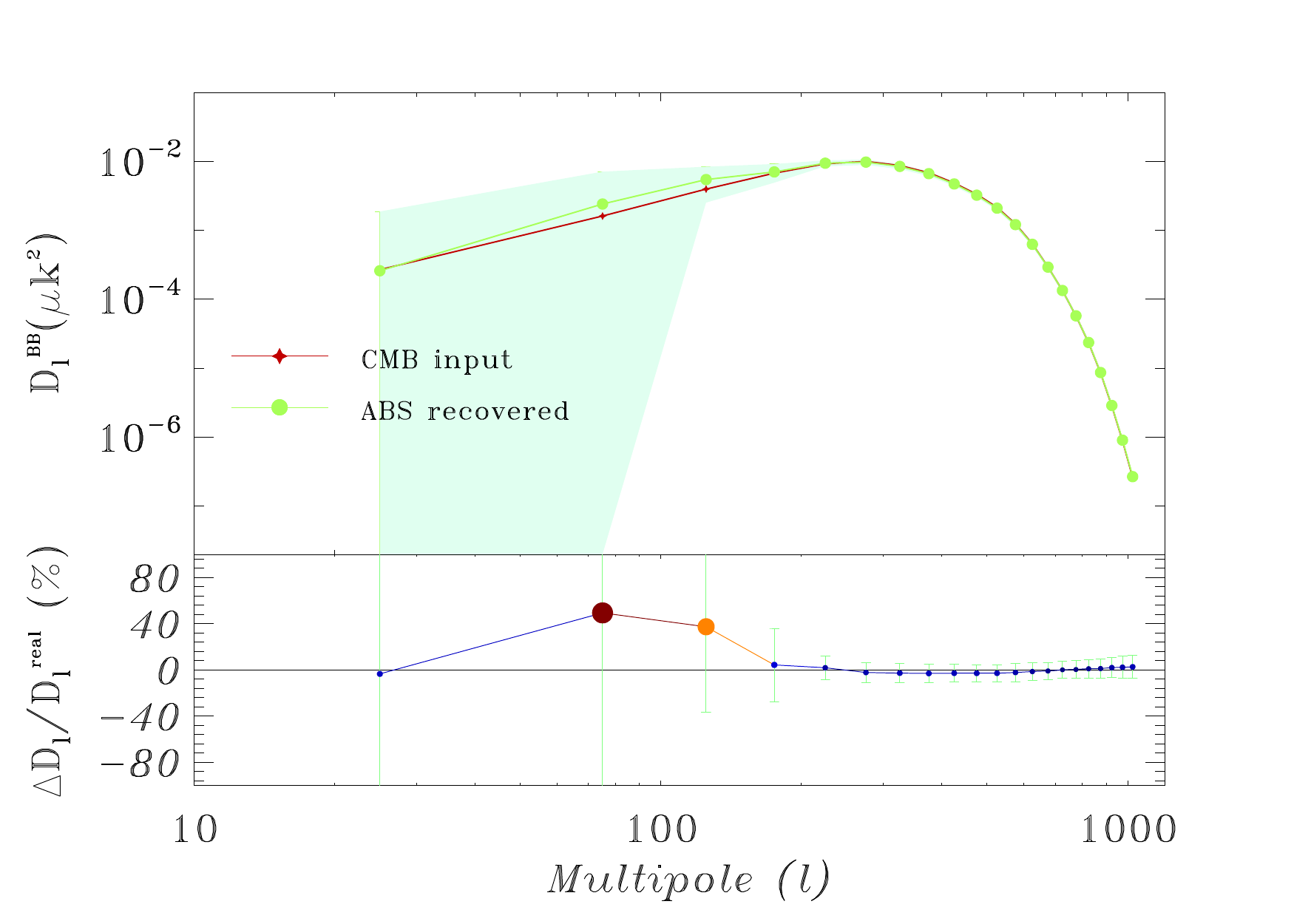}
\includegraphics[width=3.5in] {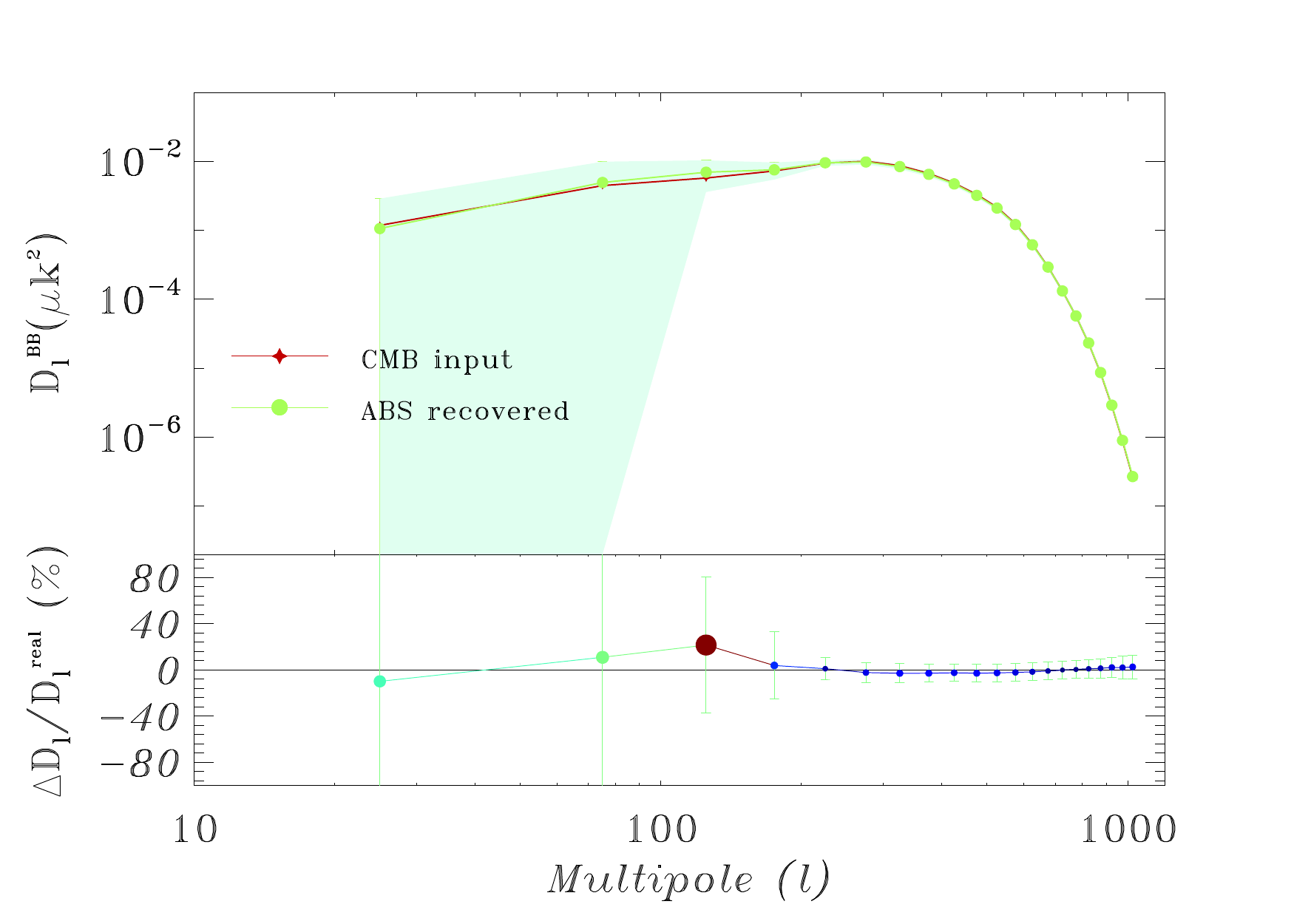}
\caption{$B$-mode power spectra obtained with ABS for a full-sky analysis with three foreground components including polarized AME, for both $r$=0 (top panel) and $r$=0.05 (lower panel). 
} 
\label{fig:full_AME} 
\end{figure}

The $B$-mode power spectrum estimation results for $r=0$ and $r=0.05$ are shown in Figure~\ref{fig:full_AME}. 
The input power spectrum rests inside the ABS recovery error bars, which are, however, big for the multipoles at low and intermediate $\ell$. Once more, the spectrum is well recovered for $\ell>170$, with a relative error $<5\%$. For $\ell<170$, this deviation reaches $50\%$ in the intermediate scales, but remains within error bars, which are large for this full-sky analysis.

\begin{figure}[!htpb]
\centering
\includegraphics[width=3.5in] {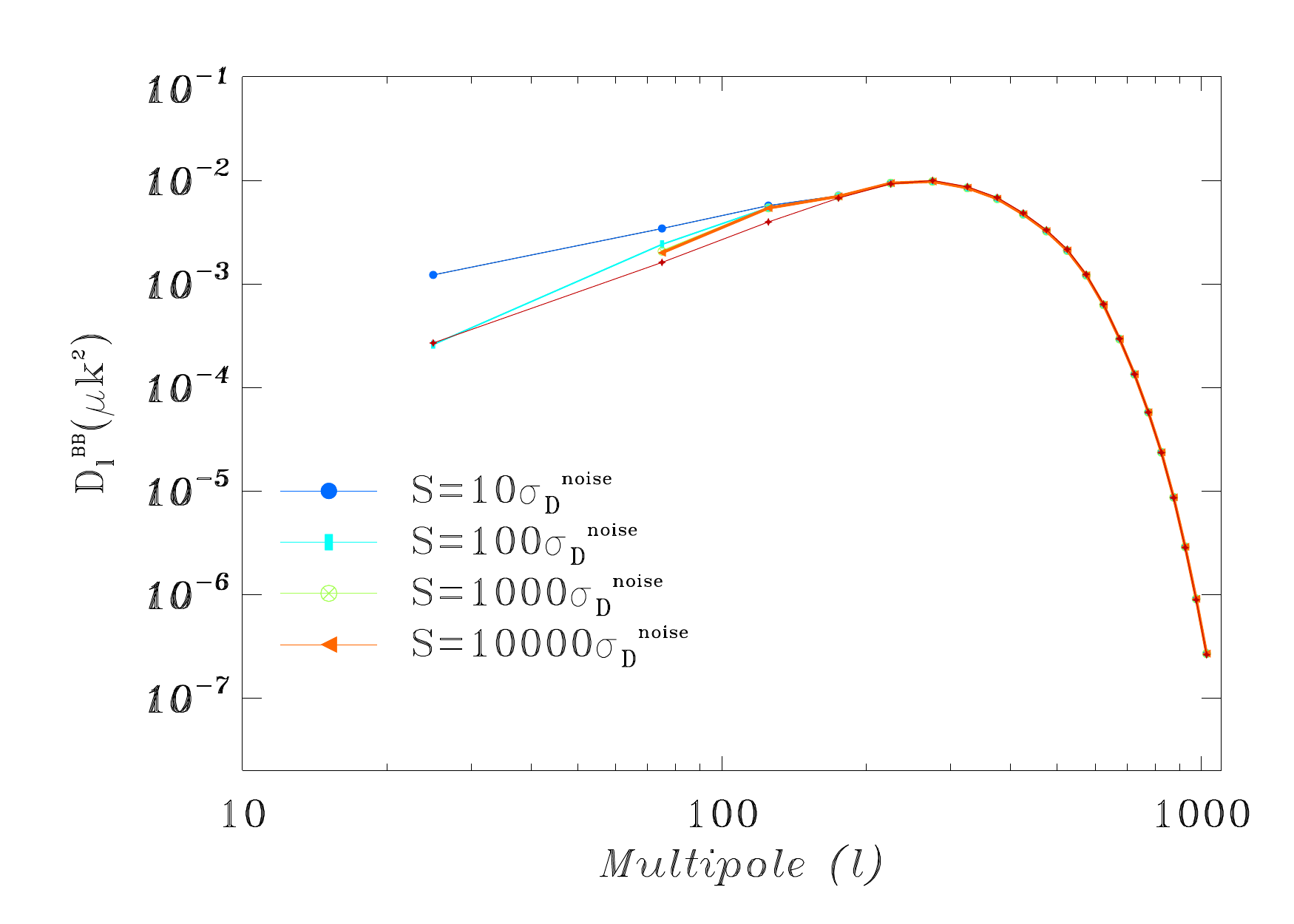}
\caption{Reconstructed $B$-mode power spectrum for $r=0$ with different values for the shifting parameter $\mathcal{S}$.
} 
\label{fig:AME-shift-parameter} 
\end{figure}

\begin{figure*}[htb]
\centering

     \mbox{
 \subfigure{
   \includegraphics[width=2.3in] {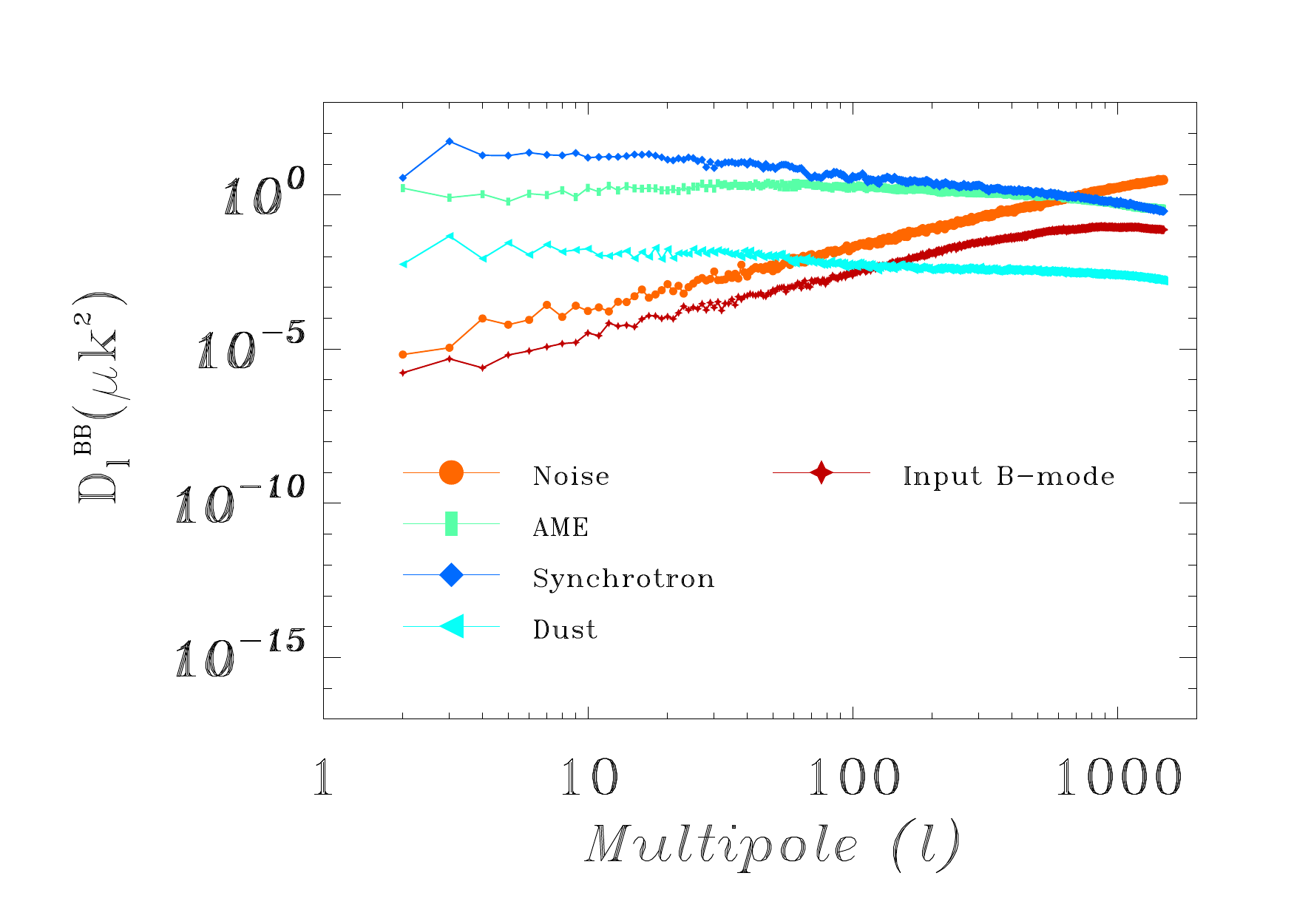}
   }

 \subfigure{
   \includegraphics[width=2.3in] {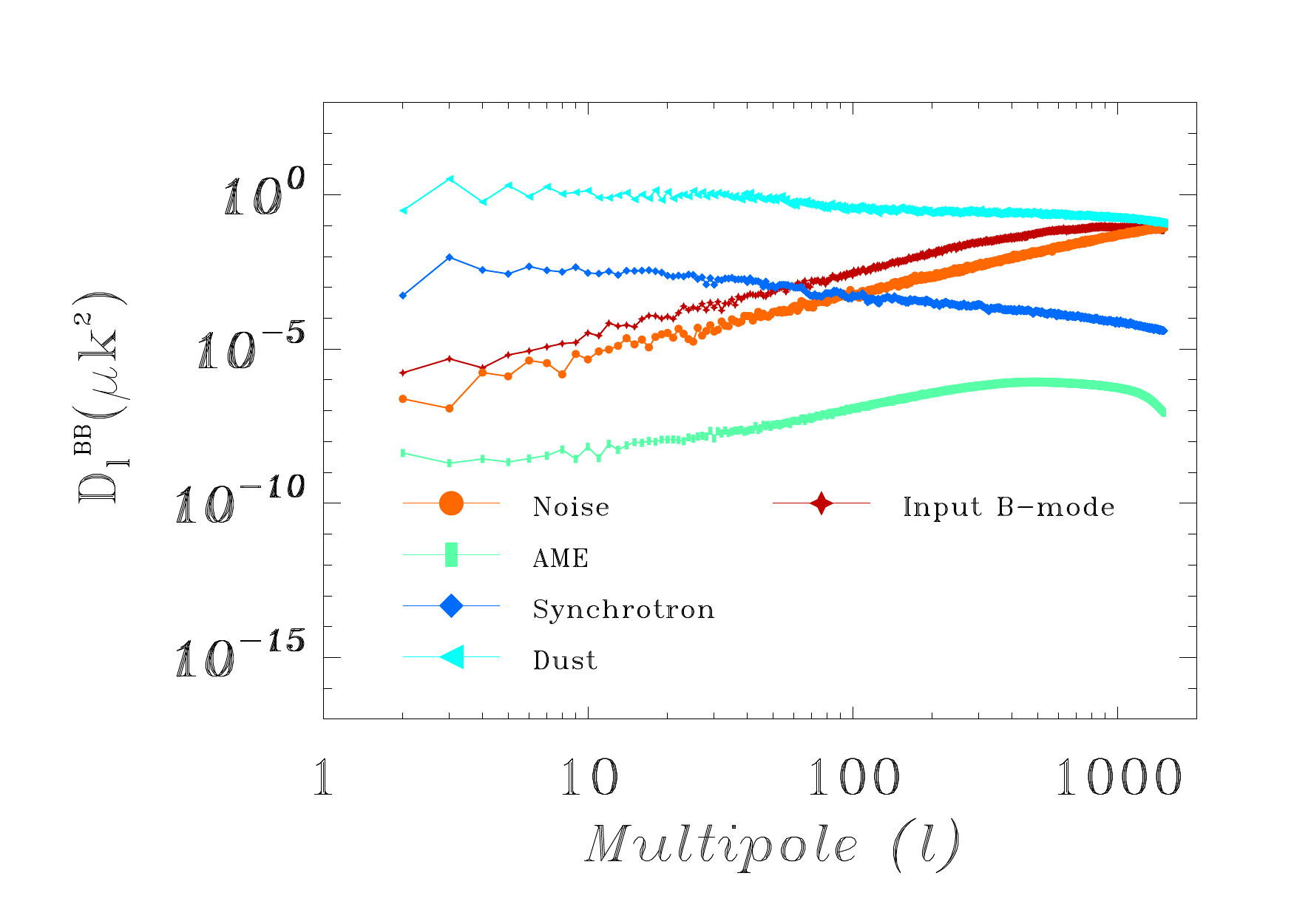}
   }

 \subfigure{
   \includegraphics[width=2.3in] {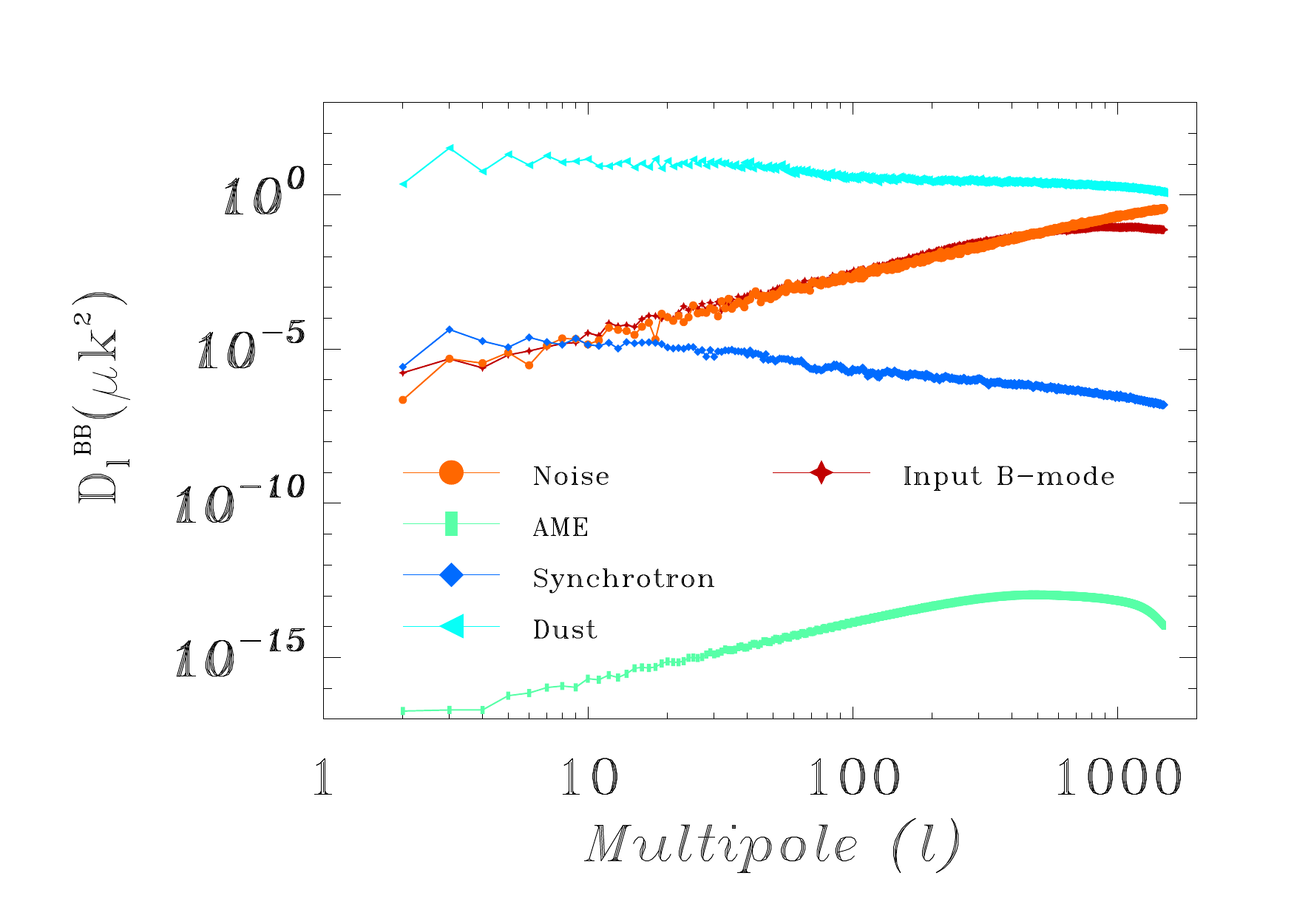}
   }
}
\caption{The  $B$-mode angular power spectra of the simulated full-sky maps (without any beam convolution), with a comparison of all the simulated components for three frequency channels: CMB, noise, synchrotron, thermal dust and AME. From left to right: 30GHz, 129GHz, and 321GHz.}
\label{fig_PS_AME}
\end{figure*}

In Figure \ref{fig:AME-shift-parameter}, we show the test for the shift parameter (for $r=0$). Here, we observe an underestimate of the $B$-mode spectrum in the first bin already for $\mathcal{S}=1000\sigma_{D}^{noise}$, and we choose as our baseline $\mathcal{S}=100\sigma_{D}^{noise}$. 


The results shown in this appendix agree with the results in Section \ref{results}, which states that the larger errors at low $\ell's$ obscure the disentanglement of the primordial $B$-mode signal from foregrounds and noise. It is also important to point out that an extra foreground increases the deviations of the ABS reconstruction in respect to the ``true" CMB for the low and intermediate multipoles. The main limitation in these results, again, comes for the very high level of foreground emission as compared to the target CMB signals. This is immediately visible in Figure~\ref{fig_PS_AME}, which shows the power spectra of the simulated foregrounds, noise, and CMB inputs for three frequency channels as an example. In addition to very strong synchrotron and dust contamination, the AME is seen to be very significant at low frequencies for the polarization level that was assumed here.

\section{Note on $E/B$ separation}
\label{app:E-B}

We now look in more detail at the impact of the performance of our $E/B$ separation method with the pseudo-$C_\ell$ estimator on the reconstructed power spectra. For this purpose, we perform 300 simulations of pure CMB signal without noise or foreground. The input power spectra used in the simulations and the Gaussian apodized mask are the same as the ones used for the partial-sky analysis of Section \ref{sect:partial_sky}. We consider, for illustration, the case of $r=0$.

\begin{figure*}[!htpb]
\centering
\includegraphics[width=0.45\textwidth] {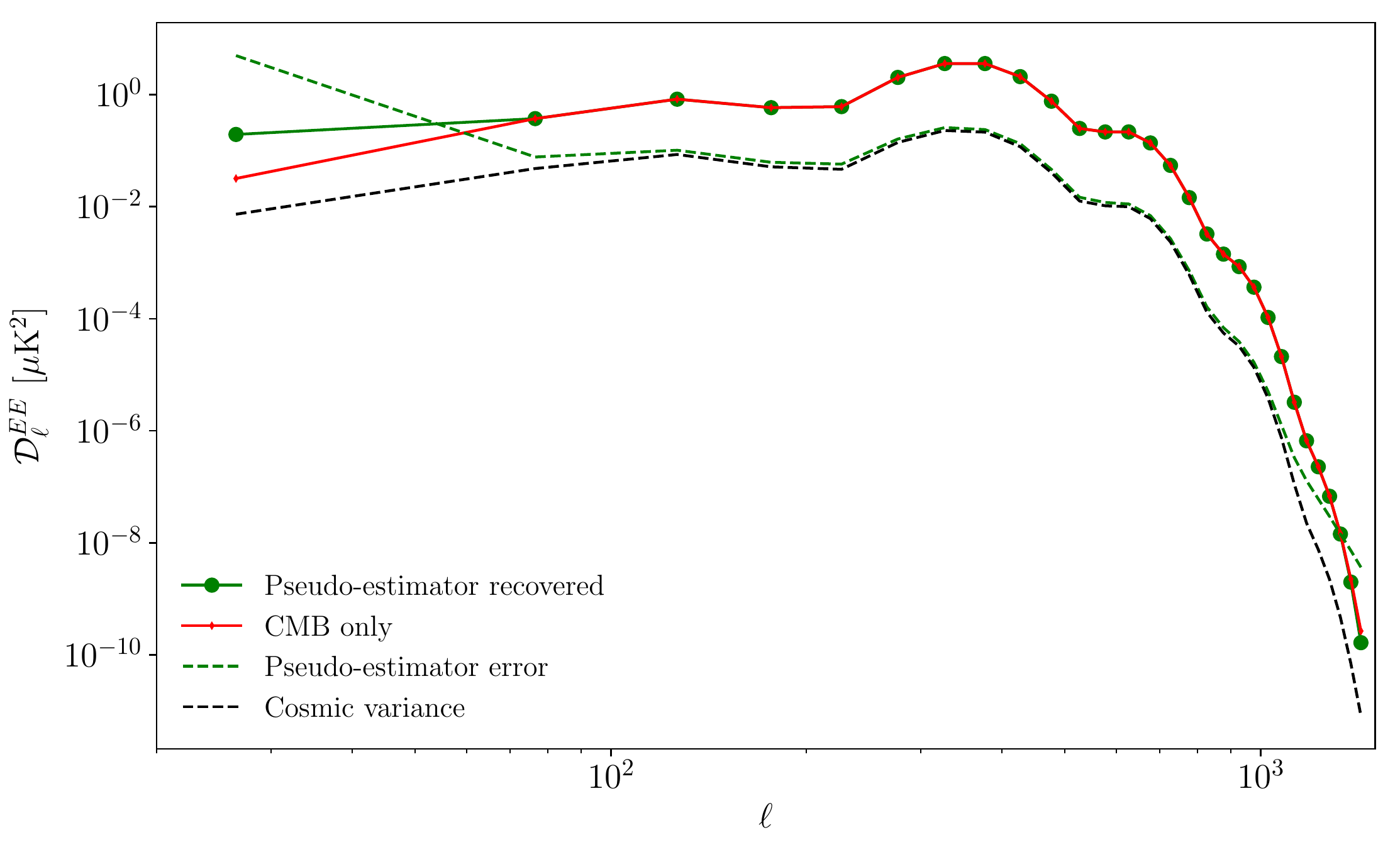}
\includegraphics[width=0.45\textwidth] {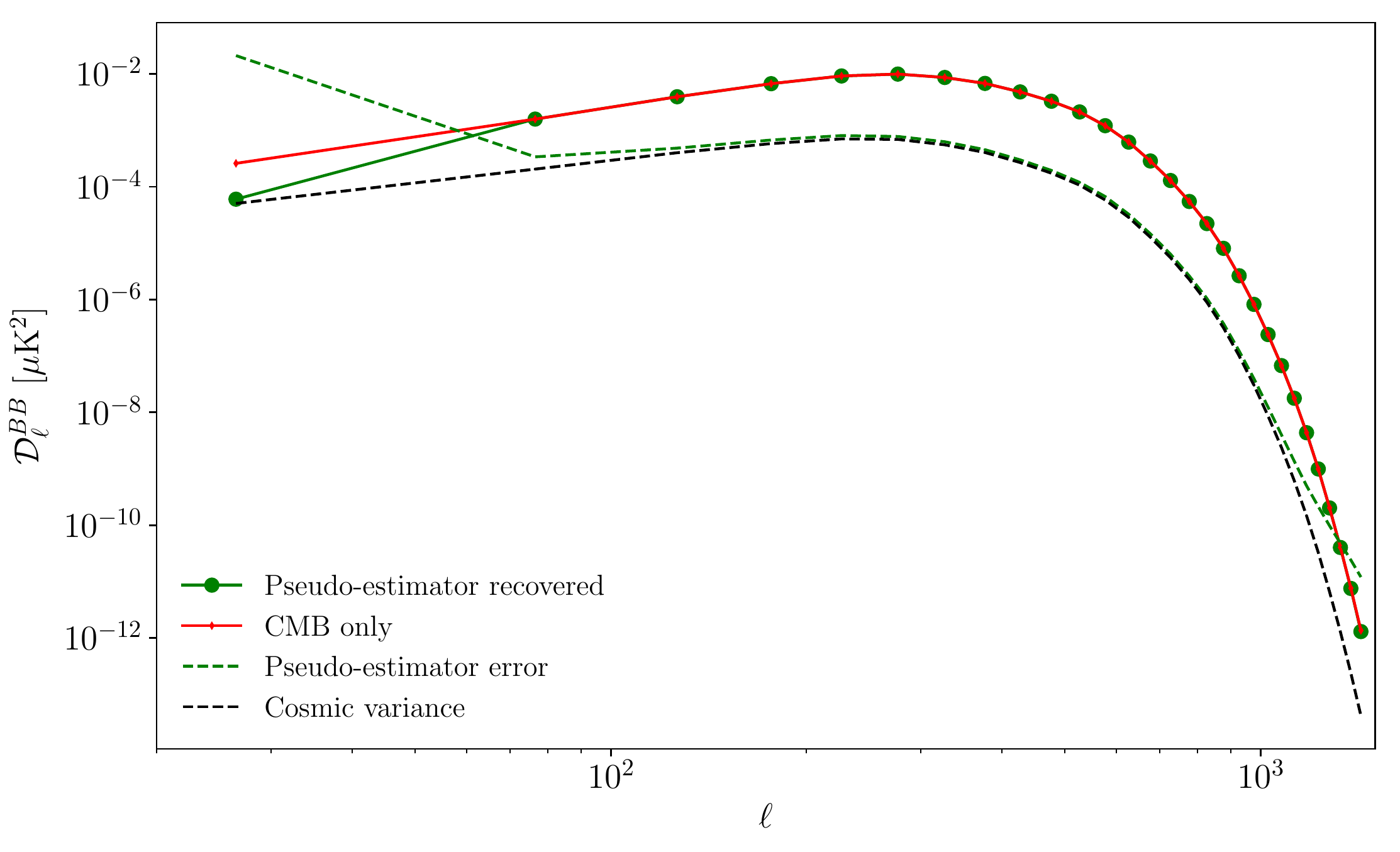}
\caption{The result for the recovered $E$-mode (left panel) and $B$-mode (right panel) spectra for 300 pure CMB realization simulations and $r=0$. The red lines shows the input CMB power spectra, the solid green lines the mean of the reconstructed power spectra. The green dashed lines represents the standard deviation of the reconstructed power spectra. The black dashed lines are displaying the cosmic variance} 
\label{EB-sep} 
\end{figure*}


The resulting $E$-mode and $B$-mode spectra are shown in Fig.~\ref{EB-sep}. While the mean reconstructed spectra above $\ell = 50$ match very well with the input power spectra, both the $E$- and $B$-mode pseudo-power spectra deviate substantially from the input spectrum.


We find that for the range $100\leq \ell \leq 1000$ the errors for both the $E$-and $B$-mode power spectra are consistent with the cosmic variance limit. This means that the pipeline is optimal in the range $100 \leq \ell \leq 1000$. But the error sharply increases below $\ell=100$ and above $\ell=1000$. It is evident from the calculations presented here that the Smith-Zaldarriaga method does not perform very well for the largest angular scales (ie., $\ell \leq 50$) with the mask we use here.

This behavior also explains the partial-sky results shown previously in Section \ref{sect:partial_sky}. The large uncertainty and deviation of the reconstructed power spectra in the first bin  are mainly due to the effects from the $E/B$ separation method. In other words, the discrepancy observed at low $\ell$s for the ABS reconstruction in partial sky would be attributed to the $E/B$ separation method. These results also suffer from similar increase in the error at high-$\ell$s, which will have a large contribution from the $E/B$ separation method itself.

 

\section{Special case}
\label{appendix:special-case}



Consider the very special hypothetical case where the sky is observed with two channels only, one of which measures a perfect CMB signal with no foregrounds and no noise, and the other one measures only white noise, with no CMB signal.
Then, we have two data sets, a pure CMB signal
\begin{equation}
    x_1(\ell,m) = s(\ell,m),
\end{equation}
and a pure noise signal
\begin{equation}
    x_2(\ell,m) = n(\ell,m).
\end{equation}
Obviously, in this simple case, the spectral estimation problem is trivially solved by disregarding the noise channel, and estimating the CMB power spectrum as
\begin{equation}
    \widehat C_\ell =  \frac{1}{2\ell+1}\sum_m x_1(\ell,m).
\end{equation}

Let's assume that we use instead the ABS prescription. Consider one specific value of $\ell$, and denote as
$\sigma_s^2 = \langle x_1^2 \rangle$ 
and $\sigma_n^2 = \langle x_2^2 \rangle$, where $\langle . \rangle$ denotes the ensemble average. 
Assuming that the noise is not correlated with the CMB signal, the ensemble average $\langle x_1x_2^* \rangle$ vanishes. The true covariance matrix of the observations is
\begin{equation}
\label{eq:true-covar}
    \mathcal{D}_{\rm obs} = \begin{pmatrix} \sigma_s^2 & 0 \\
    0 & \sigma_n^2 \end{pmatrix},
\end{equation}
which is already diagonal. After subtracting the noise covariance, eigenvectors are $\vec e_1 = (1,0)^T$ and 
$\vec e_2 = (0,1)^T$, with associated eigenvalues $\lambda_1=\sigma_s^2$ and $\lambda_2=0$. We keep only the first in the sum of Eq.~\ref{eq:abs}, and we get 
$\widehat C_\ell = \sigma_s^2$.

In practice however, ABS starts with the empirical covariance of the observations, i.e. using estimators 
\begin{equation}
\widehat \sigma_s^2 = \frac{1}{2\ell+1}\sum_m x_1^2 ,
\end{equation}
\begin{equation}
\widehat \sigma_n^2 = \frac{1}{2\ell+1}\sum_m x_2^2 ,
\end{equation}
and the cross-covariance term is 
\begin{equation}
\frac{1}{2\ell+1}\sum_m x_1x_2^* = \rho \widehat \sigma_n \widehat \sigma_s,
\end{equation}
where $\rho$ is the empirical correlation due to the finite sample size. The empirical covariance matrix is, instead of that of Eq.~\ref{eq:true-covar},
\begin{equation}
\label{eq:empirical-covar}
    \widehat {\mathcal{D}}_{\rm obs} = \begin{pmatrix} \widehat \sigma_s^2 & \rho \widehat \sigma_s \widehat \sigma_n \\
    \rho \widehat \sigma_s \widehat \sigma_n & \widehat \sigma_n^2 \end{pmatrix}.
\end{equation}
After subtracting the noise covariance (assumed to be known with negligible error), and adding the shift parameter $\mathcal{S}$, we get
\begin{equation}
\label{eq:abs-covar}
    \widehat {\mathcal{D}}_{\rm obs} = \begin{pmatrix} \widehat \sigma_s^2 + \mathcal{S} & \rho \widehat \sigma_s \widehat \sigma_n \\
    \rho \widehat \sigma_s \widehat \sigma_n & \widehat \sigma_n^2 -\sigma_n^2 \end{pmatrix}.
\end{equation}
Let $\epsilon = \rho \sigma_s\sigma_n/(\widehat \sigma_s^2 + \mathcal{S})$, and $\delta=(\widehat \sigma_n^2 -\sigma_n^2)/(\widehat \sigma_s^2 + \mathcal{S})$. For $\mathcal{S} \gg \sigma_n^2$ and $\sigma_s \sim \sigma_n$, both $\epsilon$ and $\delta$ are small, with zero mean and standard deviations
\begin{equation}
    \sigma(\epsilon) = \frac{1}{\sqrt{2\ell+1}} 
    \frac{\sigma_s \sigma_n}{(\widehat \sigma_s^2 + \mathcal{S})} \ll 1,
\end{equation}
\begin{equation}
        \sigma(\delta) = \sqrt{\frac{2}{2\ell+1}} 
    \frac{\sigma_n^2}{(\widehat \sigma_s^2 + \mathcal{S})} \ll 1.
\end{equation}
With these notations, we get 
\begin{equation}
    \widehat {\mathcal{D}}_{\rm obs} = (\widehat \sigma_s^2 + \mathcal{S}) \begin{pmatrix} 1 & \epsilon \\
    \epsilon & \delta \end{pmatrix}.
\end{equation}
The eigenvalues, and their approximations to second order in $\delta$ and $\epsilon$, are
\begin{eqnarray}
\lambda_1 & = & \frac{(\widehat \sigma_s^2 + \mathcal{S})}{2} \left[ 
1+\delta+\sqrt{(1-\delta)^2+4\epsilon^2}
\right] \nonumber  \\
& \simeq & (\widehat \sigma_s^2 + \mathcal{S})(1+\epsilon^2)
\end{eqnarray}
and
\begin{eqnarray}
\lambda_2 & = &\frac{(\widehat \sigma_s^2 + \mathcal{S})}{2} \left[ 
1+\delta-\sqrt{(1-\delta)^2+4\epsilon^2}
\right] \nonumber  \\
& \simeq & (\widehat \sigma_s^2 + \mathcal{S})(\delta-\epsilon^2).
\end{eqnarray}
The corresponding eigenvectors are
\begin{equation}
    \vec e_1 \propto \begin{pmatrix} 2\epsilon \\
    \delta-1+\sqrt{(1-\delta)^2+4\epsilon^2} 
\end{pmatrix},
\end{equation}
    and
\begin{equation}
    \vec e_2 \propto \begin{pmatrix} 2\epsilon \\
    \delta-1-\sqrt{(1-\delta)^2+4\epsilon^2} 
\end{pmatrix},
\end{equation}
which after normalization, and to second order in $\delta$ and $\epsilon$, can be approximated as
\begin{equation}
    \vec e_1 \simeq \begin{pmatrix} 1-
    \textstyle\frac{1}{2}\epsilon^2 \\
    \epsilon 
\end{pmatrix}
\end{equation}
    and
\begin{equation}
    \vec e_2 \propto \begin{pmatrix} \epsilon + \delta\epsilon \\
    -1+\textstyle\frac{1}{2}\epsilon^2 
\end{pmatrix}.
\end{equation}
Computing $G_1^2\lambda_1^{-1}$ and $G_2^2\lambda_2^{-1}$, we get
\begin{equation}
    G_1^2\lambda_1^{-1} \simeq \frac{(1-2\epsilon^2)}{(\widehat \sigma_s^2 + \mathcal{S})}
\end{equation}
\begin{equation}
    G_2^2\lambda_2^{-1} \simeq \frac{\epsilon^2 (1+\delta)^2} {(\delta-\epsilon^2)(\widehat \sigma_s^2 + \mathcal{S})}.
\end{equation}
When all goes well, $\lambda_2 \sim (\widehat \sigma_s^2 + \mathcal{S})\delta = (\widehat \sigma_n^2 -\sigma_n^2) \ll \sigma_n^2$, and the second term is neglected in the ABS solution (Eq.~\ref{eq:abs}). In that case, keeping only the first eigenvalue and eigenvector, we get (after subtraction of $\mathcal{S}$),
\begin{equation}
    \mathcal{D}^{\rm cmb} \simeq \widehat \sigma_s^2 \left( 1 +  \frac{2 \rho^2 \widehat \sigma_n^2}{(\widehat \sigma_s^2 + \mathcal{S})} \right) .
    \label{eq:case1-bias1}
\end{equation}
The second term in the parenthesis is a positive bias, which is reduced by increasing $\mathcal{S}$, and, because of the scaling of the variance of $\rho$ with $\ell$, is more important for large scales.

For the lowest $\ell$ values, it can happen that the second eigenvalue $\lambda_2 \simeq (\widehat \sigma_n^2- \sigma_n^2)$ is larger than the threshold $\lambda_{\rm cut}\sigma_n^2$. When this is the case, the second eigenvalue is included in the sum to estimate $\mathcal{D}^{\rm cmb}$. We get, instead of Eq.~\ref{eq:case1-bias1}, the solution
\begin{equation}
    \mathcal{D}^{\rm cmb} \simeq \widehat \sigma_s^2 \left( 1 -  \frac{\rho^2 \widehat \sigma_n^2}{(\widehat \sigma_n^2 - \sigma_n^2)} \right) .
\end{equation}
We now have a negative bias, proportional to $1/(2\ell+1)$, which is due to the empirical correlation between the CMB and the noise. It is the ABS equivalent of the ILC bias discussed by \citet{2009A&A...493..835D} (notably in the appendix). We note that if we had not subtracted the noise expectation value from the covariance of the observations (as is the case with the ILC), the power of the CMB map would have been 
\begin{equation}
\mathcal{D}^{\rm cmb} \simeq \widehat \sigma_s^2 \left( 1 -  \rho^2 \right) \simeq \widehat \sigma_s^2 \left( 1 - \frac{1}{2l+1} \right). 
\label{eq:ABS-ILC-bias}
\end{equation}
The multiplicative factor can be compared to that due to the ``ILC bias'', which we recall amounts to a multiplicative term of $(1-(m-1)/N_p)$, where $m$ is the number of channels and $N_p$ the number of independent modes used to compute the covariance matrix used in the ILC. With $m=2$ observation channels and with $N_p = (2\ell+1)$ independent modes used for the computation of the statistics, the ILC bias factor is exactly the factor found in Equation \ref{eq:ABS-ILC-bias}. It is interesting that thresholding the second eigenvalue out of the sum cancels this bias, and replaces it with the positive bias of Equation \ref{eq:case1-bias1}, which can, contrarily to the ILC bias, be reduced by increasing the shift parameter~$\mathcal{S}$.

While illustrative and enlightening, the naive case discussed in this appendix does not capture (by far) the full complexity of what happens when the data comprises noisy observations in several channels, contaminated by several non stationary foreground emissions. Numerical simulations, as performed in this paper, are necessary to investigate in more depth the impact of $\lambda_{\rm cut}$ and $\mathcal{S}$ on subtle biases that may impact the measurement of faint CMB polarization signals. 



\end{document}